\documentstyle[graphicx,epsffrag,psfrag,amssymb]{article}
\input{ijmpd.tex}

\begin{document}
\setlength{\textheight}{7.7truein}    
\runninghead{[Quasi-equilibrium states across Landau horizons
in effective gravity} {Quasi-equilibrium states across Landau horizons
in effective gravity}

\normalsize\textlineskip
\thispagestyle{empty}
\setcounter{page}{1}

\copyrightheading{}		

\vspace*{0.88truein}

\fpage{1}
\centerline{\bf THERMAL QUASI-EQUILIBRIUM STATES ACROSS}
\vspace*{0.035truein}
\centerline{\bf LANDAU HORIZONS}
\vspace*{0.035truein}
\centerline{\bf IN THE EFFECTIVE GRAVITY OF SUPERFLUIDS}
\vspace*{0.37truein}
\centerline{\footnotesize Uwe R. Fischer} 
\vspace*{0.015truein}
\centerline{\footnotesize\it 
Eberhard-Karls-Universit\"at T\"ubingen, Institut f\"ur Theoretische Physik }
\baselineskip=10pt
\centerline{\footnotesize\it  Auf der Morgenstelle 14, D-72076 T\"ubingen, 
Germany
}
\vspace*{10pt}
\centerline{\footnotesize Grigori E. Volovik\footnote{Present address:
Helsinki University of Technology, Low Temperature
Laboratory, P.O. Box 2200, FIN-02015 HUT, Finland.  
}}
\vspace*{0.015truein}
\centerline{\footnotesize\it L. D. Landau Institute for Theoretical Physics}
\baselineskip=10pt
\centerline{\footnotesize\it Kosygin Str. 2, 
Russian Academy of Sciences, 117940 Moscow, 
Russia}
\vspace*{0.225truein}
\publisher{(received date)}{(revised date)}
\vspace*{0.21truein}


\abstracts{
We give an account of the physical behaviour of a quasiparticle horizon due to
non-Lorentz invariant modifications of the effective space-time experienced by
the quasiparticles (``matter'') for high momenta. By introducing a
``relativistic'' conserved energy-momentum tensor, we derive quasi-equilibrium
states of the fluid across the ``Landau'' quasiparticle horizon at temperatures
well above the quantum Hawking temperature.  Nonlinear dispersion of the
quasiparticle energy spectrum is instrumental for quasiparticle communication
and exchange across the horizon. It is responsible for the establishment of the
local thermal equilibrium across the horizon with the Tolman temperature being
inhomogeneous behind the horizon.  The inhomogeneity causes relaxation of
the quasi-equilibrium states due to scattering of thermal quasiparticles, which
finally leads to a shrinking black hole horizon. This process serves as the
classical thermal counterpart of the quantum effect of Hawking radiation and
will allow for an observation of 
the properties of the horizon at temperatures well
above the Hawking temperature. We discuss the thermal entropy related to the
horizon. We find that only the first nonlinear correction to the energy
spectrum is important for the thermal properties of the horizon. They are fully
determined by an energy of order $E_{\rm Planck}(T/E_{\rm Planck})^{1/3}$,
which is well below the Planck energy scale $E_{\rm Planck}$, so that 
Planck scale physics is not involved in determining thermal quantities 
related to the horizon.
}{}{}




\vspace*{1pt}\textlineskip
\section{Introduction}

The quantum phenomenon of Hawking radiation\cite{hawkingnature} 
is of kinematical origin, and relies on the
existence of a Lorentzian signature background metric for the propagating
particle, {\it i.e.}, it depends upon the causal structure of space-time.
It does not depend on the field equations determining
the metric\cite{vissersonic} and  is, in particular,
invariant under conformal transformations of the metric\cite{kangted}.
Lorentz invariance, which is a symmetry of nature experienced
to hold on large enough scales, may not be an exact symmetry on
the smallest scales. Because a particle at an infinite distance from the
black hole horizon experiences an infinite blueshift when traced back
to the horizon, the horizon is a suitable means to
probe the high frequency structure of particle spectra: Black hole
physics is necessarily being influenced by the large
momentum modes of the quantum fields.

To trace the influence of the ultraviolet region upon  the
Hawking radiation from the black hole horizon, the condensed matter
analogue of a black hole provides an orientation
guide, with known underlying physical laws\cite{vissersonic,unruh1}.
In condensed matter,
ultraviolet deviations from ``Lorentz invariance'' naturally arise,
since the Lorentz symmetry itself is the consequence of the linear
spectrum of excitations and thus exists only in the low energy corner.
This allows for the use of physically imposed deviations from Lorentz
invariance, which are 
governed by the Planck energy scale. In a number of papers,
therefore, nonlinear dispersion of the quasiparticle spectrum has been employed
to investigate the quantum physical behaviour related to the black hole 
horizon\cite{unruh2,BHdispers,Corley,BHlaser,origin}. Below, we will find 
that in the most important
cases only the first nonlinear correction determines the physics 
of the horizon, 
since the largest energies involved, although being determined by the
Planck energy, are everywhere much smaller than the Planck scale.

Though initially the  condensed matter analogue of black hole has been
introduced in a normal liquid, it appeared that the properties of the
quantum vacuum can be incorporated properly only if one uses quantum superfluid
liquids\cite{grishated,parallel}.   The superfluid ground state moving
without friction, and an inhomogeneous superfluid velocity ${\bf v}_{\rm s}$,
provide the quantum vacuum, whose inhomogeneity serves as the gravitational
field background. The matter propagating  in the presence of this background is
represented by fermionic (in Fermi superfluids) or bosonic (in Bose
superfluids) quasiparticle excitations above the superfluid ground state.
They form the normal component of the liquid. Two fluid
hydrodynamics, introduced by Landau and Khalatnikov, incorporates
the motion of the superfluid background and excitations\cite{Khalatnikov}, 
and forms the counterpart of the Einstein equations, which
incorporate both gravity and matter.
A closer equivalence reveals itself when the quasiparticles possess
``relativistic'' properties in their low energy corner, so that the
superfluid background provides an ``acoustic'' metric\cite{vissersonic} 
of Lorentzian
signature for the low energy excitations. This occurs for example in the
case of Bogoliubov fermionic quasiparticles in $^3\!$He-A\cite{parallel}, and
phonons in superfluid $^4\!$He, as well as for excitations
in dilute Bose condensates.
The spectrum deviates
from Lorentz invariance for larger energies in a well-defined way, fixed
by the exact Bogoliubov spectrum in $^3\!$He-A, respectively
the phonon-maxon-roton spectrum in $^4\!$He.
In $^3\!$He-A, the analogy is even more attractive, since
gravity arises naturally among the other low energy collective modes of the
superfluid vacuum\cite{parallel}, in the same manner as in
Sakharov's theory of induced gravity\cite{sakharov}.

According to Landau's theory of superfluidity, the motion of the superfluid
vacuum is frictionless until a critical velocity is reached, at  which
quasiparticle creation from the vacuum becomes possible. In
superfluids with ``relativistic''-like spectrum of quasiparticles, the
Landau critical velocity equals the corresponding ``speed of
light'' relevant for quasiparticles, which is the linear slope of the
energy spectrum. For a generic profile of the superfluid velocity
${\bf v}_{\rm s}$-field, ``superluminal'' motion of the
superfluid vacuum leads to formation of a
horizon for ``relativistic'' quasiparticles. We will call such a horizon
``Landau horizon''. If the boundaries of the container are situated
far enough away from the horizon, the
quasiparticle creation predicted by Landau for superluminal flow occurs
by a quantum process at the Landau horizon analogous to
the Hawking radiation process at a gravity horizon.

The value of the Hawking temperature $T_{\rm H}=(\hbar/2\pi)\kappa_{\rm s}$ is
determined by the surface acceleration $\kappa_{\rm s}$ 
on the black hole horizon. In the
black hole analogy of fluids\cite{vissersonic,unruh1}, the corresponding relevant quantity is
the flow velocity gradient along the horizon normal,
so that for purely radial flow
$T_{\rm H}=(\hbar/2\pi)(dv_{\rm s}/dr)|_{\rm hor}$.  In
the presence of a Landau horizon in superfluids, Hawking radiation leads
to {\em quantum friction} -- the superfluid motion is decelerated due to
dissipation caused by the Hawking radiation 
process\cite{grishated,grishapainleve}.

The value of the Hawking temperature serves also as a crossover measure  
to distinguish
the region of conventional thermodynamics of the quasiparticle gas at
$T\gg T_{\rm H}$ and the quantum regime $T< T_{\rm H}$, for which dimensional
quantization of the quasiparticle energy levels becomes important. The 
Hawking temperature is typically rather small even for condensed matter
analogues of the horizon, which makes it difficult to observe quantum
effects. Therefore,  it will prove 
useful to discuss those thermal consequences of
the horizon at $T\gg T_{\rm H}$ which can be simulated by the thermal states
of quasiparticles in the presence of a flowing superfluid background. 

In the absence of a Landau horizon the true equilibrium thermal states ({\it
i.e.} those with constant temperature and without entropy production) exist for
any velocity field ${\bf v}_{\rm s}$. They correspond to Tolman's law in a
gravity field with constant Tolman temperature, cf.\cite{Tolman}, \S 129. In the presence of the Landau horizon the
situation changes: True equilibrium states no longer exist. This means
that any thermal state of superfluid liquid is dissipative in the presence of a
horizon. We  find such dissipative quasi-equilibrium thermal states across the
Landau horizon using the simplest example of superflow in 1+1 space-time
dimensions. They are characterized by constant true temperature $T$ 
outside the
horizon (Tolman's law in gravity),  and varying $T$ behind it, which satisfies
a modified Tolman law.  The physics of these states is determined by deviations
from Lorentz invariance in the ultraviolet region, in which quasiparticle
excitations can propagate superluminally. The superluminal propagation across
the horizon allows quasiparticles to exchange information and energy, and thus
ensures the continuity of the temperature across the horizon.  At $T\gg T_{\rm
H}$, the dissipation caused by quantum effects related to Hawking radiation can
be neglected, and the relaxation (and final extinction) of the black hole
occurs via entropy production behind the
horizon, where the ``Tolman temperature'' is not constant. The dissipation is
caused by the thermal conductivity and viscosity of the excitation gas (the
normal component of the liquid). 

The high energy corrections to the quasiparticle spectrum are responsible
for the processes of dissipation  in our  1+1d case.  These processes
are absent in the purely relativistic r\'egime of massless excitations, if the
so-called {\em Conformal Killing Vector} (CKV)  conditions are satisfied, see,
{\it e.g.},\cite{Zimdahl}. These CKV conditions are realized for
the 1+1d case, where 
thus dissipation can be caused only by deviations from Lorentz
invariance in the ultraviolet region. Dissipation is most pronounced in the
vicinity of the horizon due to the blueshift for quasiparticles there,  
so that the horizon region is responsible for the largest part of the entropy
production. This dissipation process 
represents the classical counterpart of the quantum process of
entropy production by  Hawking radiation from the horizon.
The same high energy corrections relevant for entropy production
determine also the entropy of the horizon itself.

In the second section to follow, we review the
properties of the fermionic quasiparticle spectrum in $^3\!$He-A and
the bosonic one in $^4\!$He. This  
comprises the influence on the energy spectrum of the
superfluid background  as an effective gravity field, the formation of the
Landau horizon and the structure of the superfluid vacuum in the presence
of a horizon, taking into account the nonlinear dispersion at higher
energy and, finally, the Hawking radiation process in the presence of 
such a dispersion.  

In the third section, using the ``acoustic''
Painlev\'e-Gullstrand metric, 
we reformulate the hydrodynamic equations for the normal
component in a ``relativistically'' invariant form, so that they
correspond to the thermodynamics of massless
1+1-dimensional 
relativistic particles moving 
in a gravitational field.  

The thermal states across the horizon are obtained as solutions of the
hydrodynamic equations on both sides of the horizon in the fourth 
section, with matching between
the states occuring due to non-Lorentzian corrections, which are 
most pronounced near the Landau horizon. 
While the equilibrium thermal state outside the horizon is
determined as usual by a {\em timelike} 
Killing vector, the thermal state behind the 
horizon satisfies the general equilibrium condition generated by 
a {\em spacelike} conformal Killing vector.   

In section five the entropy related to the
thermal states close to the horizon is calculated, and 
dissipation behind and at the horizon is discussed. 
To compare the thermal and quantum processes of the black hole relaxation, we
consider the quantum correction to the thermal energy in our flow geometry.
It has a negative sign and thus, as usually assumed, can compensate the
diverging term in the stress-energy tensor for the Hartle-Hawking state at $T=
T_{\rm H}$. However, we argue that this state nevertheless remains 
dissipative due to the  nonlinear dispersion at high energy, so that 
the black hole horizon must eventually shrink at any temperature.

\section{Spectrum and motion of quasiparticles}
\subsection{Phonons in a moving superfluid}
A suitably simple framework for our purposes is a
superfluid with massless excitations, which have a relativistic-like
spectrum at low energy. There are two classes of such systems:
bosonic and fermionic superfluids.
A bosonic superfluid is a degenerate Bose system which experiences
Bose-Einstein condensation. In the low energy limit there are only bosonic
excitations in such liquids. In the very low energy corner these are
phonons, whose energy spectrum is  $E^{\rm com}=cp$,
where $c$ is the speed of sound, and energy is measured
in the reference frame comoving with the superfluid
condensate.  This is realized in superfluid $^4$He and in
laser-manipulated, ultracold Bose-Einstein condensed gases.
At higher momentum $p$ the energy spectrum deviates from the
relativistic linear behaviour:
\begin{equation}
E^{\rm com}=cp +\gamma p^3+ \ldots \equiv cp\left(1 \pm {1\over 2}  
\left({cp \over E_{\rm Planck}}\right)^2 + \ldots \right)
\,,
\label{PhonoSpectrum}
\end{equation}
where $\gamma$ can be positive or negative. The parameter $\gamma$ is
typically determined by the analog of the  Planck energy $E_{\rm Planck}$ in
superfluids. It will be shown, for positive
$\gamma$, that while the first nonlinear correction is important for the
horizon problem, the higher order corrections are in many cases small and can be
neglected. In these cases ``transPlanckian'' physics is not involved, 
{\it i.e.}, no ``quantum gravity'' is necessary.

The excitation energy in the laboratory frame, 
in which the superfluid velocity ${\bf v}_{\rm s}$, the
momentum and the energy of quasiparticles are measured, is obtained by a
Galilean transformation which has the form
\begin{equation}
{\bf p}={\bf p}^{\rm lab}={\bf p}^{\rm com}\,, \qquad   E=
E^{\rm lab}({\bf p})=
E^{\rm com}({\bf p}) +{\bf p}\cdot{\bf v}_{\rm s}~\,,
\label{GalileanTransform}
\end{equation}
or, using (\ref{PhonoSpectrum}),
\begin{eqnarray}
\left(E-{\bf
p}\cdot{\bf v}_{\rm s}\right)^2
=c^2 p^2 +2\gamma c p^4 + \ldots \equiv c^2 p^2
\left( 1 \pm \left({c p \over E_{\rm Planck}}
\right)^2
\right) + \ldots\,
\label{EbasicPhonons}
\end{eqnarray}
This equation for the spectrum corresponds to what we would obtain for a plane
wave mode from a relativistic scalar wave equation in the WKB
approximation, modified in the comoving frame
by higher order derivatives to include nonlinear
dispersion\cite{BHdispers}.
If we neglect nonlinear corrections, we can write the energy dispersion
(\ref{EbasicPhonons})  in the form $g^{\mu\nu}p_\mu p_\nu = 0$, where
the contravariant components of the metric are
$g^{00}=-1,\,$   $g^{0i}= -v_{\rm s}^i$, $g^{ij}= c^2 \delta^{ij} -v_{\rm s}^i
v_{\rm s}^j$.
In this Lorentz invariant limit, $cp \ll E_{\rm Planck}$, the phonons become
``relativistic''  massless particles of energy $E=-p_0 $,
propagating in a (3+1)-dimensional space-time background  with
metric $g_{\mu\nu}$ of Lorentzian signature. This effective metric takes the
Painlev\'e-Gullstrand form
\begin{equation}\label{PainlevGullstrand}
ds^2= -(1-(v_{\rm s}/c)^2)dt^2 - 2c^{-2}{\bf v}_{\rm s} \cdot d {\bf r}dt   +
c^{-2}  (d{\bf r})^2 \,,
\end{equation}
with
$\sqrt {-g}=c^{-3}$.
Observe that we define the metric
coefficients such that the space-time interval has dimension of time, so that
the quasiparticle energy has the same dimension as $p_0$.
\newpage
\subsection{Landau critical velocity, ergosurface and Landau horizon}
If the superfluid velocity exceeds the Landau critical value
\begin{equation}
v_L={\rm min}{E^{\rm com}({\bf p})\over p}\,,
\label{LandauVelocity1}
\end{equation}
the energy $E = E^{\rm lab}({\bf p})$ of some excitations, as  measured in the
laboratory frame, becomes negative. This allows for excitations to be
nucleated from the vacuum. For a superfluid velocity field which is
stationary in the laboratory frame, the surface $v_{\rm s}({\bf r})=v_L$,
which bounds the region where quasiparticles can have negative
energy, the {\em ergoregion}, is called the {\em ergosurface}.

If the dispersion bends upwards, $\gamma>0$, the Landau critical velocity
coincides with the ``speed of light'', $v_L=c$, so that the ergosurface
is determined by $v_{\rm s}({\bf r})=c$.
In the Lorentz invariant limit, this corresponds
to $g_{00}({\bf r})=0$, which is just the definition of the
ergoregion in gravity.  In the case of radial flow (zero axial component of
${\bf v}_{\rm s}$) of the
superfluid vacuum towards the origin, the ergosurface becomes the horizon
in the Lorentz invariant limit, and the region inside the horizon
simulates a black hole for low energy phonons. Strictly
speaking this is not a true horizon for phonons: Due to the nonlinear
dispersion, their group velocity $v_g=dE^{\rm com}/d p
=c+3\gamma p^2>c$, and thus the high energy quasiparticles are allowed to
leave the black hole region. It is, hence, a horizon only for quasiparticles
living exclusively in the very low energy corner, which are not aware
of the possibility of ``superluminal'' motion.

In the case of negative dispersion, $\gamma<0$, the group velocity
$v_g=dE^{\rm com}/d p <c$. In such superfluids the relativistic ergosurface
$v_{\rm s}({\bf r})=c$ does not coincide with the true ergosurface, which is
determined by $v_{\rm s}({\bf r})=v_L< c$.  In superfluid $^4\!$He, 
where $\gamma<0$, 
the Landau velocity is related to the roton part of the spectrum,
and is about four times less than $c$.\footnote{The phonon dispersion 
in superfluid $^4\!$He actually first has
a slight upturn of phonon dispersion (and hence $\gamma >0$) for small
phonon momenta. If we assign $\gamma <0 $, it is thus meant that the 
dispersion {\em after} this slight upturn, for intermediate momenta, 
 bends down to the roton minimum.}\,  For inward radial flow, the
ergosphere occurs at $v_{\rm s}(r)=v_L<c$, while the inner surface $v_{\rm
s}(r)=c$ still marks the horizon, in contrast to relativistically invariant
systems, for which ergosurface and horizon coincide for purely radial flow.

\subsection{Painlev\'e-Gullstrand {\it versus} Schwarzschild
metric in effective gravity}
In our context, it appears worthwhile to point out that in the effective theory
of gravity, which occurs in condensed matter systems in the form that the
primary quantity is the contravariant metric tensor
$g^{\mu\nu}$ describing the energy spectrum,
two seemingly equivalent representations of the black hole metric,
in terms of either the Schwarzschild or the
Painlev\'e-Gullstrand line elements, are in fact not
equivalent in terms of the required stability of the underlying
superfluid vacuum. Let us construct an isotropic Galilean superfluid in
3+1 dimensions, in such a way that the low energy spectrum of the   Bogoliubov
quasiparticles in the  frame comoving with the superfluid is given by
\begin{equation}
(E^{\rm com})^2=c^2p^2~,
\label{IsotropicSupewrfluid}
\end{equation}
where $c$ is the parameter which plays
the part of the speed of light. Then, for spherically symmetric
velocity profiles $v_{\rm s}(r)=\mp c \sqrt{r_{\rm S}/r}$,
 the Painlev\'e-Gullstrand line elements  are
obtained in the laboratory frame as:
\begin{equation}\label{gullstrand}
ds^2= -\left(1-{r_{\rm S}\over r}\right)dt^2 + 2c^{-1}\sqrt{r_{\rm S}\over r}drdt
+ c^{-2}  dr^2 +(r^2/c^2)\,d\Omega^2 \,,
\end{equation}
\begin{equation}\label{gullstrand2}
ds^2= -\left(1-{r_{\rm S}\over r}\right)dt^2 - 2c^{-1}\sqrt{r_{\rm S}\over r}drdt
+ c^{-2}  dr^2 +(r^2/c^2)\,d\Omega^2 \,,
\end{equation}
where $r_{\rm S}$ is the Schwarzschild radius.
The Painlev\'e-Gullstrand line elements in
(\ref{gullstrand},\ref{gullstrand2}) describe black hole and
white hole analogues  (see {\it e.g.} ref.\cite{vissersonic}, 
a pedagogical review of the Painlev\'e-Gullstrand metric
is contained in\cite{Martel}).  The ``surface gravity''  at the Schwarzschild
radius is, therein, 
$\kappa_{\rm s}  =dv_{\rm s}/dr|_{r_{\rm S}} = c /2r_{\rm S}$,  and the
 Hawking temperature which follows $T_{\rm H}=\hbar\kappa_{\rm s}/2\pi$.

An ``equivalent'' representation of the black or white hole metric is given
by the Schwarzschild line element, which in terms of the same superfluid
velocity reads
\begin{equation}
 ds^2=-\left(1- v_{\rm s}^2/c^2\right)d\tilde t^2+{dr^2\over c^2-
v_{\rm s}^2} +(r^2/c^2)\,d\Omega^2\,.
\label{Schwarzschild}
\end{equation}
 They are related by the following coordinate transformation:
\begin{equation}
 \tilde t(r,t)=t +   \left({2\over v_{\rm s}(r)} + {\rm ln}~ {1- v_{\rm s}(r)\over
1+v_{\rm s}(r)}\right)\, ,\qquad d\tilde t=dt +{v_{\rm s}\over 1-v_{\rm s}^2}dr.
\label{Transformation}
\end{equation}
From the point of view of the effective gravity in superfluids, 
(\ref{gullstrand}) and (\ref{gullstrand2}) describe the dynamics of
quasiparticles propagating in the  superfluid vacuum, moving with superfluid
velocity $\mp|v_{\rm s}(r)|$, directed to the origin and away from the origin, 
respectively. The space-time $(t,{\bf r})$ 
is the absolute space-time of the laboratory frame, {\it i.e.}, as 
measured by the external experimentalist living in the real world of the
laboratory.  The squared 
energy spectrum of the quasiparticles,  determined by the
contravariant components of the metric 
(\ref{gullstrand},\ref{gullstrand2}), is obtained to be  
\begin{equation}
(E -v_{\rm s}(r)p_r)^2 =  c^2p^2\,,
\label{SpectrumInSuperflow1}
\end{equation}
or, solving for $E$, 
\begin{equation}
 E =v_{\rm s}(r)p_r \pm  cp\,.
\label{SpectrumInSuperflow2}
\end{equation}
The time $\tilde t$ is the time as measured by the
``inner'' observer at ``infinity'' (far from the hole). 
The ``inner'' means
that this observer ``lives'' in the superfluid background and makes use
of ``relativistic'' massless quasiparticles 
(phonons or other excitations) to synchronize clocks. 
The inner observer at some point $R\gg 1$ sends  
a quasiparticle pulse at the moment $t_1$ which arrives at point $r$ of 
$t=t_1+\int_{r}^R {dr'}/|v_-|$ absolute (laboratory) time, where $v_+$
and $v_-$ are absolute (laboratory) velocities of radially propagating
quasiparticles, moving outwards and inwards respectively: 
\begin{equation}
v_\pm ={dr\over dt}={dE\over dp_r}=\pm 1 +v_{\rm s}(r)\,.
\label{RadialVelocity}
\end{equation}
Since from the point of view of the inner observer the speed of light 
(the propagation speed of quasiparticles) 
is an invariant quantity, and   does not depend on
the direction of propagation, for him or her the moment of arrival 
of the pulse to $r$ is not $t$
but  $\tilde t =(t_1+t_2)/2$, where
$t_2$ is the time when the pulse reflected from $r$ returns to the observer at
$R$. Since $t_2-t_1=\int_{r}^R{dr'}/|v_-|+\int_{r}^R{dr'}/|v_+|$, one obtains
for the time measured by an inner observer 
\begin{eqnarray}
 \tilde t(r,t)& = &{t_1+t_2\over 2}=t +  {1\over 2} \left(\int^R_r{dr'\over
v_+}+
\int^R_r{dr'\over v_-}\right)\nonumber\\
& = &t +   \left({2\over v_{\rm s}(r)} + {\rm ln}~ {1-
v_{\rm s}(r)\over 1+v_{\rm s}(r)}\right)-   \left({2\over v_{\rm s}(R)} + {\rm ln}~ {1-
v_{\rm s}(R)\over 1+v_{\rm s}(R)}\right)  \,,
\label{InnerTime}
\end{eqnarray}
which is just (\ref{Transformation}), up to a constant shift.

At $r=r_{\rm S}=1$ one has an event horizon,
where the superfluid velocity crosses the "speed of light"  $c\equiv 1$.
If the fluid moves towards the origin, that is 
$v_{\rm s}(r)<0$, this velocity field reproduces the horizon of the black hole
(the so-called sonic black hole\cite{unruh1,unruh2}): 
Since the velocity of the fluid behind horizon
exceeds the speed $c$ of the propagation of quasiparticles with respect to
the fluid, the low energy quasiparticles are trapped within the horizon.
If the fluid moves  from the origin outwards, 
$v_{\rm s}(r)>0$, the velocity field reproduces a white hole horizon.

In the (necessarily) 
complete physical space-time of the laboratory, an external observer
can detect quasiparticles  radially propagating into (but not out of) the
black hole,  or out of (but not into) the white hole. The energy spectrum
of the quasiparticles remains to be  well determined both outside and
inside the horizon.  Quasiparticles cross the black hole horizon with the
absolute velocity
$v_-= -1-v_{\rm s}=-2$, {\it i.e.} with {\em twice} the
 speed of light: $r(t) = 1  -
2(t-t_0)$. In the case of a white hole horizon one has $r(t) = 1  +
2(t-t_0)$. In contrast, from the point of view of the inner observer
the  horizon cannot be reached and crossed: the horizon can be approached
only asymptotically in infinite time: $r(\tilde
t)=1+(r_0-1)\exp(-\tilde t)$.  Such a disability of the local observer, who
lives in the curved world of superfluid vacuum,  happens because he or she is
limited in his or her observations by the ``speed of light'', so that the
coordinate frame he or she uses is seriously limited in the presence of a 
horizon and becomes incomplete.

The Schwarzschild metric (\ref{Schwarzschild}) 
naturally arises for the inner observer, if the
Painlev\'e-Gullstrand metric is an effective metric for quasiparticles in
superfluids,  but not vice versa. It 
can in principle arise  as an effective metric;
however, in the presence of a horizon such a metric indicates an
instability of the underlying medium. To obtain  a line element
of the Schwarzschild metric as an effective metric for quasiparticles,  the
quasiparticle energy dispersion in the laboratory frame has to be
\begin{equation}
E^2=c^2\left(1-{r_{\rm S}\over r}\right)^2 p_r^2+c^2\left(1-{r_{\rm S}\over
r}\right) p_\perp^2~.
\label{SpectrumInSchwarzschild}
\end{equation}
In the presence of a horizon such a spectrum has sections of the
transverse momentum $p_\perp$ having $E^2<0$. The consequence  of an imaginary
frequency of excitations signals the instability of the superfluid vacuum, if
this vacuum  exhibits the Schwarzschild metric as an effective metric for
excitations:  Quasiparticle perturbations may
grow exponentially without bound in laboratory (Killing) time, like 
$\exp[t |\Im (E)|]$, destroying the superfluid vacuum.  Nothing of this kind
happens in the case of the Painlev\'e-Gullstrand line element,
for which the quasiparticle energy is real even behind the horizon.
The main difference between  Painlev\'e-Gullstrand and Schwarzschild
metrics as effective metrics is thus as follows.
The first metric leads to the slow
process of quasiparticle radiation from the vacuum at the horizon
(Hawking radiation), whereas the second one indicates a crucial 
instability of the vacuum  behind the horizon.

In general relativity, it is usually 
assumed that the two metrics can be converted
into each other by the coordinate transformation in (\ref{Transformation}).
In condensed matter, the coordinate transformation
leading from one metric to another is not that innocent
if an event horizon is present.  The reason why the
physical behaviour implied by the choice of metric  representation
changes drastically, is that the
transformation between the two line elements, $t\rightarrow t +\int^r
dr'~v/(c^2-v^2)$, is singular on the horizon, and thus can be applied
only  to a part of space-time.  In condensed matter, only those effective
metrics are physical which are determined everywhere in the real physical
space-time and come from a physically reasonable energy spectrum of
quasiparticles.  The two representations of the ``same'' metric cannot, then, 
be strictly equivalent metrics, and we have different classes of equivalence,
which cannot be transformed to each other by coordinate
transformations which are regular everywhere.  
Painlev\'e-Gullstrand metrics for black and white holes
are determined everywhere, but belong to two different classes.  The
transition between these two metrics occurs  via the singular
transformation
$t\rightarrow t +2\int^r dr'~v/(c^2-v^2)$, or via the Schwarzschild line
element, which is prohibited in condensed matter physics, as explained
above,    since it is pathological in the presence of a horizon. It is not
determined in the whole space-time, and is singular at the horizon.

Of importance is the fact that in the effective theory,
 there is no need for an
additional extension of the space-time to make it geodesically complete. This is
because the effective space-time is {\em always} incomplete (open), since it
exists only in the low energy "relativistic" corner,
 and quasiparticles escape this space-time to a nonrelativistic domain   
when their energy increases beyond the relativistic
linear approximation r\'egime\cite{grishated}.

\subsection{Bogoliubov fermions in Fermi superfluids}
Another class of quasirelativistic systems is provided by Fermi superfluids
with point nodes in the spectrum, the physically realised example in an
earth laboratory being the superfluid $^3\!$He-A,
and possibly chiral $p$-wave superconductors\cite{Rice}.
The fermionic quasiparticles in $^3$He-A are described by a Nambu
space Hamiltonian which, in the locally comoving frame, has the form
\begin{equation}
\hat H_A = {\bf \tau}_3 \epsilon({\bf p}) + {\Delta_A\over p_F}[{\bf
\tau}_1 p_x +
{\bf \tau}_2 p_y]\,,
\qquad \epsilon({\bf p})\approx {p^2 -p_F^2\over 2m^*}\approx
v_F(p-p_F)\,.
\label{Nambu}
\end{equation}
Here, $\Delta_A$
represents the amplitude of the gap,
 $p_F$ and $v_F=p_F/m^*$ are Fermi momentum
and Fermi velocity respectively. We assume that the
orbital momentum vector
$\hat{\bf l}$ is oriented along the $z$ axis and that
the absolute value of the momentum
$p=|{\bf p}|$  is restricted by $p-p_F\ll p_F$.

The Bogoliubov    spectrum of   fermions in the comoving frame
is obtained by squaring (\ref{Nambu})
\begin{equation}
(E^{\rm com})^2  = \hat H_A^2=
 \left({p^2}/{2m^*}-{p_F^2}/{2m^*}\right)^2
             +c_\perp^2(p_x^2+p_y^2)\,.
\label{Esquared}
\end{equation}
Here $c_\perp\equiv\Delta_A/p_F\ll v_F$ (the slope of the gap at the node),
plays the role of the ``velocity of
light'' in the direction perpendicular to $\hat{\bf l}$.

If the superfluid  velocity is restricted to lie in the plane perpendicular to
$\hat{\bf l}$ and does not depend on $z$, the momentum projection $p_z$ is
a conserved quantity, and one obtains a 2+1d system with
the laboratory frame energies
\begin{eqnarray}
\left(E-{\bf
p}_\perp\cdot{\bf v}_{\rm s}\right)^2
=\frac{\left(p_\perp^2-\tilde p_F^2\right)^2}{4m_*^2}
+c_\perp^2 p_\perp^2\,,\qquad \tilde p_F^2\equiv p_F^2-p_z^2\,.
\label{Ebasic}
\end{eqnarray}

In the very low energy r\'egime, {\it i.e.}, if ${\bf p}$ is chosen
to be close to $p_F\hat{\bf e}_z$, and in addition  $p^2_\perp
=p_x^2+p_y^2\ll m_*^2 c_\perp^2$, we can write the energy dispersion
(\ref{Ebasic})  in a Lorentz invariant form with (2+1)-dimensional
metric (neglecting the quasiparticle motion in the $z$ direction). This metric
again takes the Painlev\'e-Gullstrand form
\begin{equation}
ds^2= -(1-(v_{\rm s}/c_\perp)^2)dt^2 - 2c_\perp^{-2}{\bf v}_{\rm s}
\cdot d{\bf r}dt   +
c_\perp^{-2}  dr^2 \,,
\end{equation}
with
$\sqrt {-g}=c_\perp^{-2}$.

The Bogoliubov spectrum, say, for $p_z= p_F$, $\tilde p_F= 0$ reads
\begin{equation}
E =
{\bf v}_{\rm s}\cdot{\bf p}_\perp \pm
\sqrt{\left(\frac{p_\perp^2}{2m_*}\right)^2
+ c_\perp^2 p_\perp^2}
\,.
\label{MasslessEbranches}
\end{equation}
Squared, this has the same form as for phonons with positive dispersion in
(\ref{EbasicPhonons}), with the ``Planck'' energy $E_{\rm
Planck}=2m_*c_\perp^2$. This shows that the form of spectrum with cubic
dispersion is generic for any system which has ``relativistic''
quasiparticles in the low energy corner. Further on, we shall consider the
positive dispersion, $\gamma>0$.

\subsection{Physical vacua across the horizon}
\begin{figure}[!!!t]
\input{transpsfrag}
\begin{center}\includegraphics[width=0.7\textwidth]{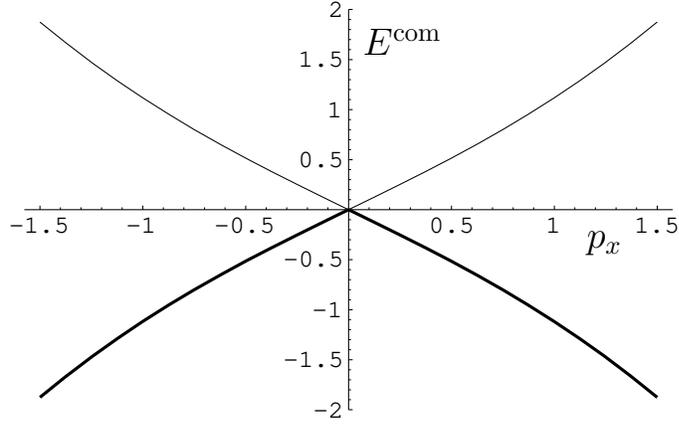}
\bigskip
\caption[EkvInside]
    {Energy spectrum in comoving frame. The occupied negative energy levels
designated by thick lines form the comoving frame vacuum.}
\label{EkComoving}
\end{center}
\end{figure}

\begin{figure}[!!!t]
\input{transpsfrag}
\begin{center}\includegraphics[width=0.7\textwidth]{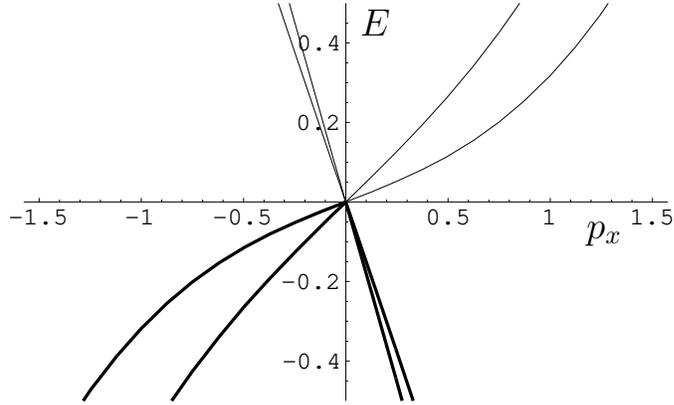}
\bigskip
\caption[EkvOutside]
    {Spectrum in the laboratory frame outside the
``horizon'' at two points in space
where  $v_{\rm s} =-0.5$ and $v_{\rm s} =-0.8$.
}
\label{EkvOutside}
\end{center}
\end{figure}

\begin{figure}[!!!t]
\input{transpsfrag}
\begin{center}\includegraphics[width=0.7\textwidth]{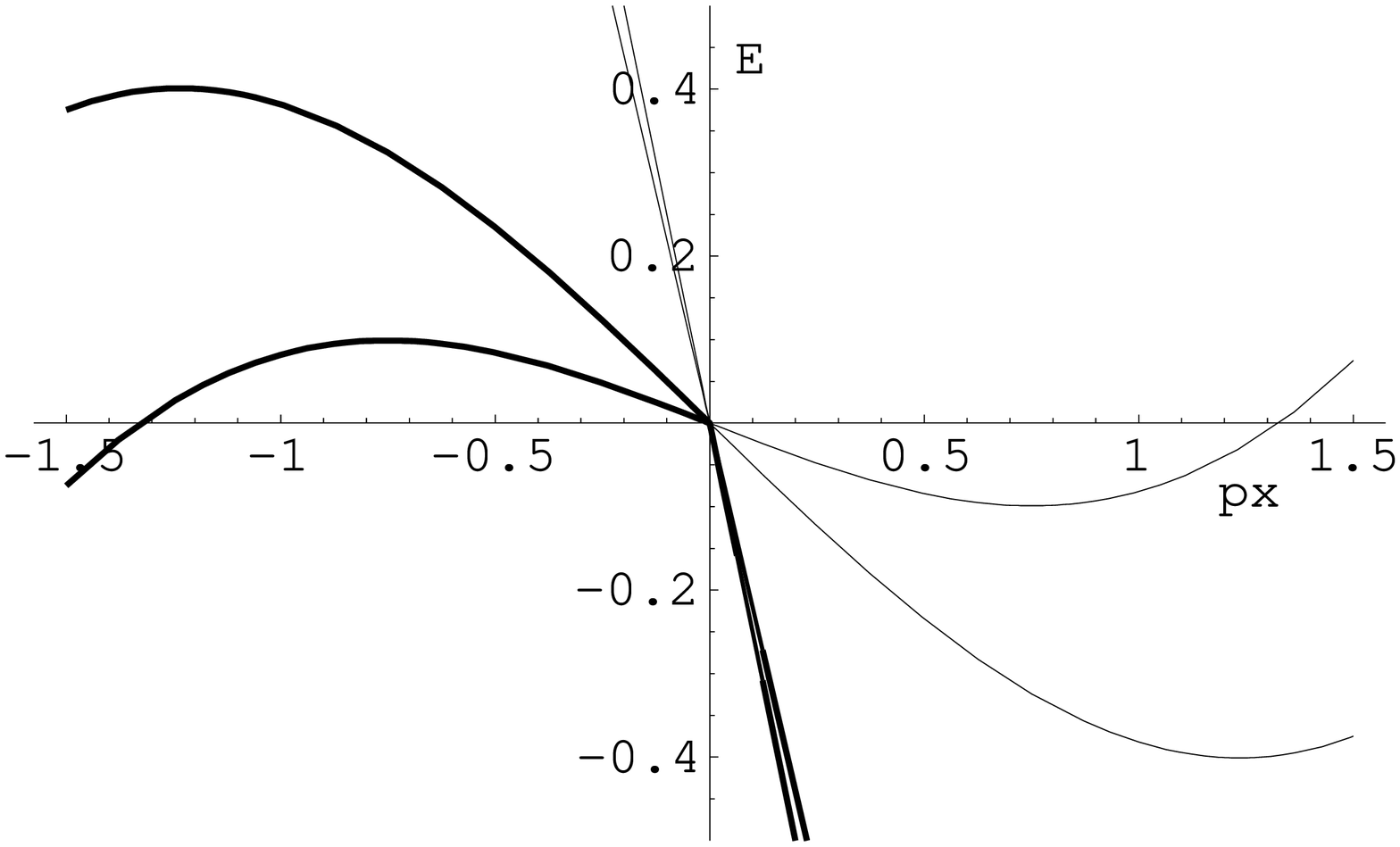}
\caption[EkvInside]
    {Spectrum in the laboratory frame inside the
``Landau horizon'' at two points where  $v_{\rm s} =-1.2$ and $v_{\rm s}
=-1.5$.
}
\label{EkvInside}
\end{center}
\end{figure}

Close to the horizon the horizon surface can be considered to be
flat, so that the radial flow becomes a one-dimensional flow of superfluid
along the normal to the horizon (along the $x$ coordinate direction),
this flow depending only on $x$ (${\bf v}_{\rm s}=v_{\rm
s}(x)\hat{\bf x}$,
with $v_{\rm s}<0$).  Since there is no dependence on
$y$ and $z$, the   momentum projections
$p_z$ and $p_y$ of quasiparticles are fixed and one obtains
(1+1)-dimensional
motion of quasiparticles whose   ``mass'' and ``speed
of light'' are determined by $p_z$ and $p_y$.
Figures \ref{EkComoving}-\ref{EkvInside} display the energy spectrum for
the case of zero ``mass'', which means $p_y=p_z=0$ for the Bose condensate,
and $p_y=0$, $p_z=\pm p_F$ for $^3$He-A. We measure $v_{\rm s}$ in units of
$c_\perp$; 
$E$  and $p_x$ in units of
$m_*c_\perp^2=\Delta_A(c_\perp/v_F)$ and $m_*c_\perp=\Delta_A/v_F$, which
represent the crossover ``Planck energy''  and the crossover ``Planck
momentum'' to the ``nonrelativistic'' domain, respectively.  Normalised
quantities in what follows will be designated by a tilde, save for $k\equiv
p_x/m_*c_\perp$ and any velocities, which are understood to be scaled by
$c_\perp$ (or $c$ in the case of the 
Bose condensate), in the normalised relation:
\begin{equation}
\tilde E =
v_{\rm s} k \pm \sqrt{k^2+ \frac{k^4}{4} }
\,.
\label{MasslessEbranchesNormalized}
\end{equation}
The position of the horizon is determined by $v_{\rm s}(x_h)=-1$. We remind
the reader that,
if the nonlinear dispersion of quasiparticles is positive, this
is not a true horizon, since the group velocity in the comoving frame
exceeds the ``speed of light''
\begin{equation}
|v^{\rm com}_g| = {1+ \frac{k^2}{2}\over \sqrt{1+ \frac{k^2}{4} }}>1
\, ,
\label{ComovingVelocityNormalized}
\end{equation}
and consequently high energy quasiparticles can propagate across the horizon.

The horizon is understood to be a  ``Landau horizon'' --
the position in space where the Landau criterium is first violated,
{\it i.e.},  where the superfluid velocity
reaches the minimum of the comoving frame group velocity:
\begin{equation}
v_{\rm s}(x_h)=-{\rm min}~|v^{\rm com}_g| = - 1
\, .
\label{LandauVelocity2}
\end{equation}
\noindent Let us first discuss 
the vacuum for the Fermi system, which corresponds to
filling of the negative energy states. The vacuum as viewed in
the {\em comoving} frame corresponds to the filling of states with negative
root in (\ref{MasslessEbranches}) (thick lines in Fig.\ref{EkComoving}).
Outside the horizon, where $|v_{\rm s}|<1$ (Fig.\ref{EkvOutside}), the
vacuum in
the laboratory frame coincides with the vacuum in the comoving frame:  In
both systems the negative energy levels are occupied.
This, however, does not happen inside the horizon, where $|v_{\rm s}|>1$
(Fig.\ref{EkvInside}), and some states, which have negative
energy in the comoving frame, have positive energy in the laboratory
frame.  In gravity, the state corresponding to the vacuum in the
comoving frame is the state empty of particles as seen by the
``free fall'' observer (who moves with superfluid velocity
$v_{\rm s}$). The vacuum in the laboratory frame corresponds to the Boulware
vacuum in gravity: It is empty for the observer at infinity. In the purely
relativistic r\'egime in 1+1 dimension the vacuum behind the horizon,
determined in the comoving frame,
is obtained if the (negative) momentum of upper index,
$p^1=(1-v_{\rm s}^2)p_1 -v_{\rm s}p_0=(1-v_{\rm s}^2)p +v_{\rm s}E$, is
used as the Hamiltonian, $H=-p^1$, corresponding to the interchange of
timelike and spacelike coordinates behind a horizon.
The states with positive
$p^1$ are occupied in this vacuum, and
$p^1=-i[(1-v_{\rm s}^2)\partial_x -v_{\rm s}\partial_t]$ is a conformal
Killing vector behind the horizon.
The same  Hamiltonian $H=-p^1$ determines thermal equilibrium
(section 4.2.).
However, any deviation from the
linear relativistic r\'egime leads to dissipation in this vacuum or the
thermal equilibrium state.

Close to the horizon, the velocity profile can be linearized:
$v_{\rm s}(x)=-1 +(\kappa_{\rm s}/c_\perp)\delta x \equiv -1+  \tilde x$,
with $\delta x =x-x_h  \ll c_\perp/\kappa_{\rm s}$.
Since in the above considerations
we assumed that $\tilde E\ll 1$ and  $ \tilde x
\ll 1$,  only small $k$ are relevant and we can approximately write
\begin{equation}
k^3/8+k  \tilde x \simeq \tilde E
\,. \label{3roots}
\end{equation}
The same cubic dispersion  has been discussed in\cite{BHdispers}.  From 
equation (\ref{3roots}) it 
follows that the characteristic scales  for
$k$, $\tilde E^{\rm com}$ and $\tilde x$ are $\tilde E^{1/3}$, $\tilde
E^{1/3}$  and $\tilde E^{2/3}$, respectively. In dimensionful units these
are
\begin{eqnarray} 
E^{\rm com}=cp \sim E_{\rm Planck}^{2/3} E^{1/3} \ll E_{\rm Planck}~,
\label{CharacteristicEnergyScale}\\
|x-x_h|
\sim {c\over \kappa_{\rm s}} \left({E\over E_{\rm Planck}}\right)^{2/3}  \sim l_{\rm
Planck} \left({E\over T_{\rm H}}\right)^{2/3} \left( {E_{\rm Planck}\over
T_{\rm H}}\right)^{1/3}\gg l_{\rm
Planck}  ~~.
\label{CharacteristicLengthScale}
\end{eqnarray} 
The characteristic energy, although highly blue-shifted compared with the
laboratory energy $E$, is nevertheless much smaller than the  Planck
energy scale. Correspondingly, the characteristic region of the shell
(membrane) in the vicinity of the horizon, though very narrow, is
nevertheless large compared to the Planck length scale (the
quasiparticle energy in the laboratory frame is determined either by the 
Hawking temperature, or by the temperature of the heat bath, 
in both cases $E\geq T_{\rm H}$). 
Consequently, the higher order corrections to the nonlinear energy
spectrum are negligibly small, so that ``transPlanckian'' physics is not
involved save for its providing the parameter $\gamma$, which can be
considered as a phenomenological parameter in the effective low energy
theory.

\subsection{Wave function}

If the initial state of the superfluid is the
vacuum in the comoving frame, then in the presence of a horizon this state
represents a false vacuum: It possesses large positive energy in the
laboratory frame, so that it has to decay. If
there is an interaction with the walls of the container,
the relaxation occurs due to
fast processes of the dynamics of the 
superfluid order parameter. However, if the
interaction with  walls is screened, or the wall is very far, the comoving
inner observer loses the reference frame imposed by the walls. Now the
velocity field itself is what the inner
observer can observe. If the flow is homogeneous, there is no evidence for him
or her that the superfluid is moving.  Thus the only information comes from the
inhomogeneity of the velocity field, which establishes a special frame, in
which the velocity profile is time independent.  This means that if a 
horizon is present, its relaxation is determined by the
gradient of the superfluid velocity. 
One example is quantum Hawking radiation,
with the Hawking temperature determined by the gradient of 
the velocity at the horizon.
Another example is 
the classical thermal state in the presence of an horizon, 
whose relaxation is provided by the hydrodynamic entropy 
production induced by the velocity gradient.

\begin{figure}[hbt]
\begin{center}
\input{transpsfrag}
{\includegraphics[width=0.83\textwidth]{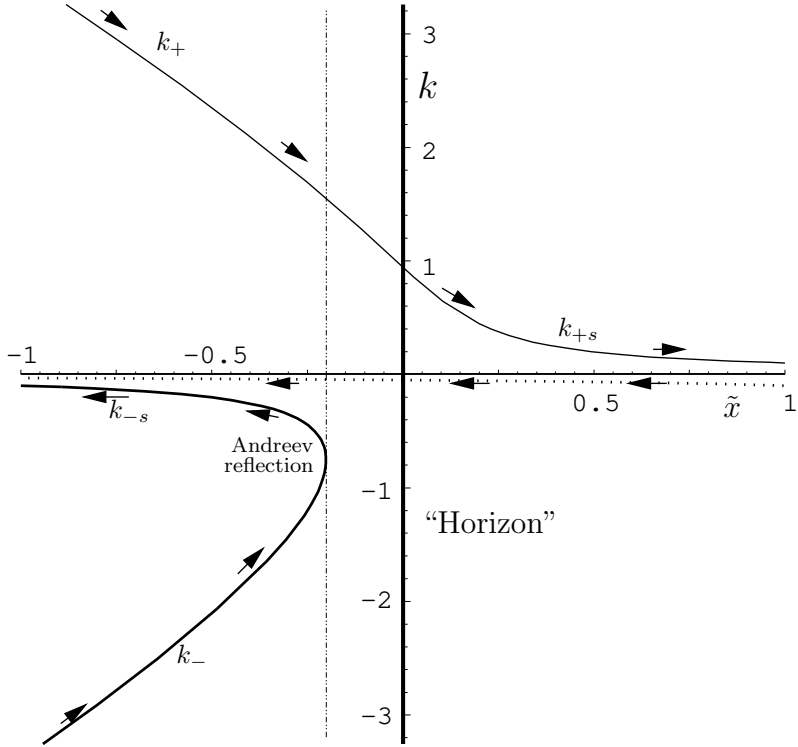}}
\caption[xkHbothE0dot1]{Quasiparticle trajectories
in the flow $v_{\rm s}= -1+\tilde x$, for $\tilde E=0.1$.
The notations are the same as those in\cite{BHdispers}: $k_{\pm}(x)$ are
part of the branches with large positive and negative momenta and
$k_{\pm s}$ are part of the branches with small positive and negative momenta.
The arrows indicate the direction of the laboratory
frame group velocity of the quasiparticle. The horizon is identical with
the $k$
axis. Thick lines for the quasiparticle trajectory
show that the states with momenta $k_{-s}$ and $k_{-}$ are
occupied in the comoving (free fall) vacuum. The dash-dotted line indicates
the closest approach distance of the branch of large negative momenta to the
horizon, at which the negative momentum
$k_-$ part of the wave coming from negative
$\tilde x$ gets Andreev reflected (and the positive momentum $k_{+s}$ part is
transmitted). The ingoing branch of very small negative momenta
is shown by the dotted line.}
\label{xkHbothE0dot1}
\end{center}
\end{figure}

Let us first recall the quantum relaxation. Using the energy $\tilde E (k)$ of
(\ref{3roots}) as a Hamiltonian, and canonically quantizing,
\begin{equation}
[ \tilde x~,k]=i\tilde\hbar\,,\qquad\tilde\hbar=\hbar\kappa_{\rm s}/(m_*c^2_\perp)\,,
\end{equation}
we write the stationary Schr\"odinger
equation in the momentum representation as
($   \tilde x=i \tilde\hbar\partial_k$):
\begin{equation}
\left[\frac{k^3}8 +\tilde\hbar
\left(k\partial_k+ \frac12\right)\right]\psi (k)
= \tilde E\psi (k)\,.
\end{equation}
This may be solved by using the ansatz
$\psi(k)=A(k)\exp[i\Phi(k)]$ to give
\begin{equation}
\psi (k) = 
\exp\left[-\frac12\ln k+i \tilde\hbar^{-1}
(-\tilde E\ln k + k^3/24)\right]\,.
\end{equation}
The wave function in $\tilde x$ space is, consequently,
\begin{eqnarray}
\psi ( \tilde x) &  = &
\int_{\cal C} dk~ 
\exp\left[-\frac12\ln k 
+i\tilde\hbar^{-1}\left(k  \tilde x
-\tilde E\ln k + k^3/24
\right)\right] \label{kintegralapprox} \\
 & = & \int_{\cal C} dk~ 
\exp\left[-\frac12\ln k+i\tilde\hbar^{-1}\left(
k  \tilde x -\int\tilde x(k)d
k\right)\right]\label{kintegral}
\,.
\end{eqnarray}
In the stationary phase (semiclassical WKB)
approximation of $ \tilde E\gg \tilde\hbar$,
the integral (\ref{kintegral}) will be
dominated by three saddle point solutions, which are
the three roots $k(\tilde E)$ of
(\ref{3roots}) for given $\tilde E$.
Further on, we consider $\tilde E>0$.
For given positive $\tilde x$ (outside the horizon),
there is only one real solution, $k_{+s}$
(we use the notation of\cite{BHdispers}).
For negative $ \tilde x$ (inside the horizon) there are three such
solutions,
$k_{-s}$, $k_+$ and $k_{-}$, provided that $ \tilde x$ is not too small (cf.
Fig.\ref{xkHbothE0dot1}). The  additional ingoing
branch in this Figure (dotted line), close to the
$x$ axis, is obtained if the full initial energy spectrum of
(\ref{MasslessEbranches}) is used. This branch has comparatively high
energy for given momentum 
and hence is not relevant for the present discussion.

Far enough from the horizon,  {\it i.e.} at $1\gg
\tilde x \gg \tilde E^{2/3}$,  we have from (\ref{3roots})
$k_{\pm s}\simeq   \tilde E/  \tilde x$,
and  $k_{\pm}\simeq \pm 2\sqrt {2 | \tilde x|}$. The branches $k_{\pm
s}=   \tilde E/  \tilde x$ exist in the linear dispersion case. At the
horizon they approach $-\infty$ and
$+\infty$, respectively. Nonlinear corrections to the energy
spectrum become important when the momentum scale $\tilde
E^{1/3}$ is  reached, and these low momentum
branches match with the high energy branches
$k_{\pm}$, which exist due to nonlinearity of the spectrum.
The choice of the path $\cal C$  in the complex $k$ plane, determining
the wave function, was discussed in detail in\cite{Corley}.

Depletion of the false vacuum occurs by tunneling from the trajectory
$k_-\cdots k_{-s}$ to the outgoing mode trajectory
 $k_+ \cdots k_{+s}$. In the WKB
limit of $2\pi\tilde E/\tilde\hbar  \gg 1$, the high energy mode $k_-$
is Andreev reflected\cite{andreev}
to $k_{-s}$, with a small probability of transmission
to the outgoing mode. The transmission coefficient is exponentially small and
corresponds to thermal radiation at the Hawking temperature:
\begin{equation}
{\cal T}\simeq\exp[- 2\pi\tilde E/\tilde\hbar]= \exp[-E/T_{\rm H}]
\,,\label{transcoeff}
\end{equation}
where the Hawking temperature is given by the equality $\tilde T_{\rm H}=
\tilde\hbar/2\pi$. The Hawking temperature
establishes a characteristic scale for the laboratory energy
$\tilde E\sim\tilde T_{\rm H}$ (or $ E\sim T_{\rm H}$) and thus the scales for
$p$,
$E^{\rm com}$ in
(\ref{CharacteristicEnergyScale}) and for the thickness of the shell
$x_0=|x-x_h|$ in (\ref{CharacteristicLengthScale}). In dimensionless units  one
has
$k\sim\tilde  E^{\rm com}\sim \tilde T_{\rm H}^{1/3}$, $\tilde x_0\sim
\tilde T_{\rm H}^{2/3}$.

The process of Andreev reflection occurs inside
the horizon for our superluminal dispersion, whereas it is situated outside
for subluminal dispersion\cite{origin}.  In both cases the Hawking
temperature does not depend on the dispersion.
The point where the wave packet is turning back in
co-ordinate space occurs where $d \tilde x /dk=0$,
which is given by solving the equation
$k_0^3/4=-\tilde E\sqrt{1+k_0^2/4}$. For small $k_0$, there results a
value $k_0\simeq -(4{\tilde E})^{1/3}=- 1.59 {\tilde E}^{1/3}$,
implying $\tilde x_0=\tilde x(k_0) =
-(4^{-1/3}+4^{2/3}/8){\tilde E}^{2/3} =-0.945{\tilde E}^{2/3} $.

\subsection{Nonlinear velocity profiles}

An asymptotic form of the velocity profile
$v_{\rm s}= 
f(\tilde x)$,
where $f(\tilde x)$ is some function approaching zero
(or a constant) 
at large positive $\tilde x$, and linearity
at small $\delta\tilde x$ around the horizon,
{\it e.g.} $f(\tilde x)=-1+\tanh
[\delta\tilde x])$, gives us the location of the quasiparticle in
terms of its wave number as
\begin{equation}
\tilde x (k)= f^{-1}\left( 
\tilde E/k -\sqrt{1+{ k}^2/4}\right)\,.
\end{equation}
We then obtain a Hamiltonian
\begin{equation}
 \tilde H = {\tilde E}^{\rm com}({ k}) 
+ \{
{ k} f( {\tilde x})
\}\,,\label{generalH}
\end{equation}
the curly brackets specifying that the appropriate operator ordering
be taken for the expression contained within.

\begin{figure}[!!!t]
\begin{center}
\input{transpsfrag}
\includegraphics[width=\textwidth]{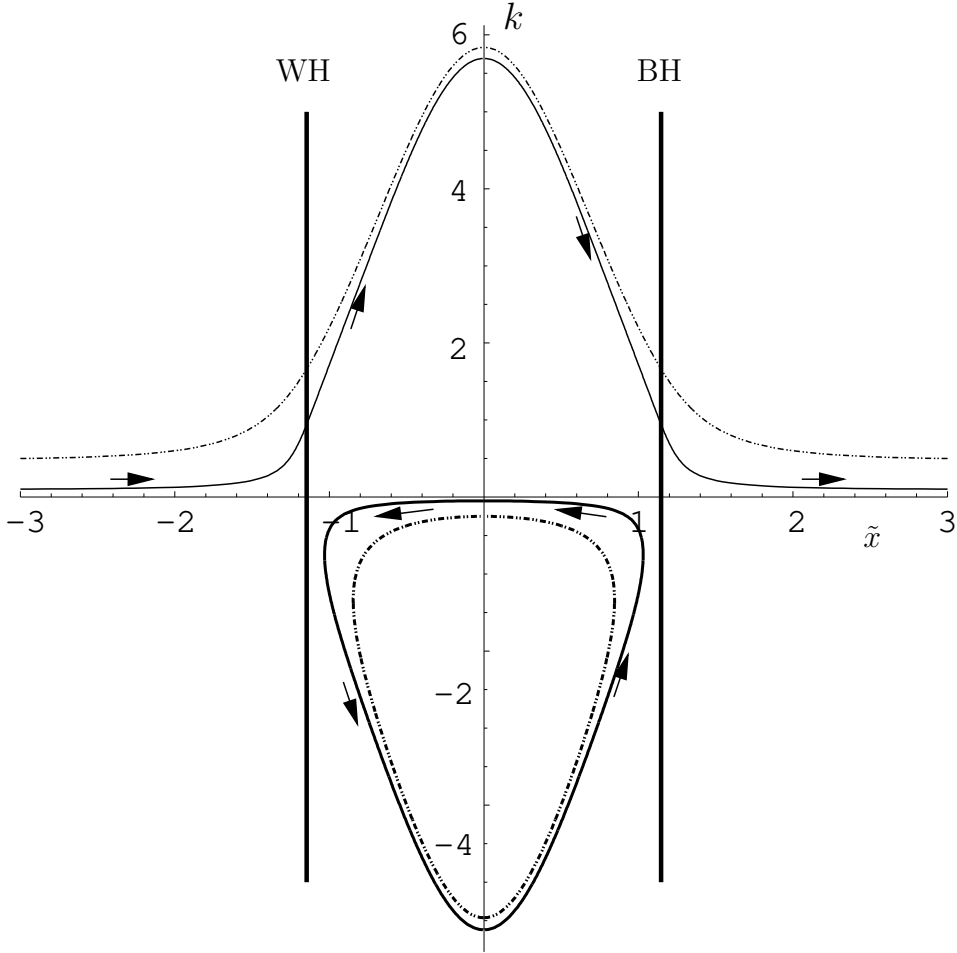}
\caption[returnnewthickbeta3mu0]
    {\label{BHWH} Large momentum quasiparticle
phase space trajectories for 
the two (fixed) energies $\tilde E=0.1$ and $\tilde E=0.5$
(dashed), in the velocity profile
$v_{\rm s}=-\beta /\cosh^2[\tilde x/\alpha]$ ($\beta > 1$), with $\alpha=1$,
$\beta =3$. The black and white hole horizons where $v_{\rm s}=-1$ are located
at $\tilde x_h=\pm \alpha \cosh^{-1}(\sqrt{\beta})$, the upper (lower)
sign valid for the black hole (BH) and white hole (WH), respectively.
The horizon surface velocity gradient
$\kappa_{\rm s}= \pm (2/\alpha)\sqrt{1-\beta^{-1}}$.
The quasiparticle states for negative
momenta are bound states, where the values of the turning points at
large $k$ are in leading order $k_t=\pm 2\beta$ for large $\beta$.}
\label{returnnewthickbeta3mu0}
\end{center}
\end{figure}

The influence of a  particularly interesting flow profile on
quasiparticle motion is depicted in
Fig.\ref{returnnewthickbeta3mu0}.
The profile is uni-directional in the negative $x$ direction, and is
developing locally superluminal velocities in a depression of the
(symmetric) profile. It creates
a characteristic combination of white hole and black hole
(cf. \cite{grishated,grishapainleve} for $^3$He-A 
and \cite{BoseCondensate}  for a Bose condensate). Specifically, there are
bound quasiparticle states inside the horizon, which occur because of the
combined effects of Andreev reflection, caused by nonlinear dispersion, and
the shape of the velocity profile.

\newpage

\section{Dynamics of the normal component (``matter'')}
Now we consider the classical temperature region $T\gg T_{\rm H}$, where the
quantum processes of Hawking radiation can be neglected. Still, 
the behaviour of matter in the presence of the horizon is nontrivial and
its description requires deviations from Lorentz invariance due to the
blueshift in the vicinity of the horizon.

\subsection{``Relativistic''
energy-momentum tensor of two fluid hydrodynamics}
The basic conservation equations
of standard nonrelativistic
Landau two fluid hydrodynamics (as expounded in detail 
in\cite{Khalatnikov} \S 8), are those of {\em total} momentum and energy
conservation. Energy and momentum can be exchanged between the two
subsystems of quasiparticles and superfluid vacuum in a way similar to the
exchange of energy and momentum between matter and the gravitational field.
Moreover, in the low energy corner this exchange can be described in
general relativistic form. The energy-momentum tensor  derived from
standard conservation equations of two fluid 
hydrodynamics\cite{Khalatnikov} can be represented as\cite{EMomtensor}:
\begin{equation}
\sqrt{-g} T^\mu{}_\nu=\sum_{\bf p} f v_g^\mu p_\nu\,,\qquad
v_g^\mu v_{g\mu} = -1 +\frac1{c^2}
\frac{\partial E^{\rm com}}{\partial p_i}
\frac{\partial E^{\rm com}}{\partial p_i}\,,
\end{equation}
where $f$ is the quasiparticle distribution function and   $\sum_{\bf p}
\equiv  \sum_{s}\int d^Dp /(2\pi\hbar)^D$ designates summation over spin
as well as momentum. The group four-velocity is defined as
\begin{eqnarray}
v_g^i& =& \frac{\partial E}{\partial p_i}\,,\quad v_g^0=1\,,\quad
v_{gi}= \frac1{c^2} \frac{\partial E^{\rm com}}{\partial p_i}
=\frac1{c^2} (v_g^{\rm com})^ i\,,\nonumber\\
v_{g0} & = & -\left(
1+\frac{1}{c^2}v_{\rm s}^i \frac{\partial E^{\rm com}}{\partial
p_i}\right)\,.
\end{eqnarray}
Space-Time  indices $\mu,\nu,\ldots$ are throughout
assumed to be raised and lowered by the Painlev\'e-Gullstrand metric.
The group four-velocity is null in the relativistic domain of the
spectrum only: $v_g^\mu v_{g\mu}=0$ if $E^{\rm com}=p$.
Here and in what follows, we again set $c=c_\perp=1$ (scale velocities
with $c$), except where the ``speed of light'' is explicitly 
indicated for clarity.

The energy-momentum tensor thus defined satisfies the
covariant conservation law equation
\begin{equation}
T^\mu{}_{\nu;\mu}={1\over \sqrt{-g}} {\partial (T^\mu_{\nu}\sqrt{-g})\over
\partial x^{\mu}} -{1\over 2}{\partial g_{\alpha\beta}\over
\partial x^{\mu}}T^{\alpha\beta} = 0
 \,.
\end{equation}
As pointed out by Landau and Lifshitz\cite{LandauLifshitz2}, this equation
does not generally express any conservation law whatever. The same is true for
quasiparticles in superfluids. This equation does not mean that energy and
momentum of quasiparticles are necessarily 
conserved: In the presence of an inhomogeneous
condensate the second term describes the energy and momentum  exchange  between
quasiparticles (matter) and superfluid vacuum (gravity field)
\begin{equation}
(\sqrt{-g}T^\mu{}_\nu)_{,\mu}= P_i \partial_\nu v_{\rm s}^i
+ 
\sum_{\bf p} f \partial_\nu E^{\rm com}
 \,,
\end{equation}
where  ${\bf
P}=\sum_{\bf p} {\bf p} f({\bf r},{\bf p})$ is the quasiparticle
momentum density. The last term appears only if the ``speed of light''
depends on space-time. Further on we neglect this term.

In a one-dimensional situation, 
 and assuming that there is no temporal dependence
of superfluid velocity, respectively if it is permitted that it is 
neglected, we obtain an energy conservation law for quasiparticles in the
form
\begin{equation}
\frac\partial{\partial t}(\sqrt{-g} T^0{}_0)
+ \frac\partial{\partial x}(\sqrt{-g} T^x{}_0)
=0\,,\label{Econserv}
\end{equation}
which tells us that in any time independent
situation, the energy flux $\sqrt{-g} T^x{}_0$ is necessarily constant in
space. The momentum conservation in $x$
direction reads 
\begin{equation}
\frac\partial{\partial t}(\sqrt{-g} T^0{}_x) + \frac\partial{\partial x}
(\sqrt{-g}T^x{}_x)
=-P_x \partial_x v_{\rm s} =-{\kappa\over v_{\rm s}} \,  T^0{}_x\sqrt{-g}
\,.  \label{Mconserv}
\end{equation}
It contains   the velocity gradient as a source term, which corresponds
to the ``gravity''
$\kappa=\partial_x (v_{\rm s}^2/2)$, where $v_{\rm s}^2/2$ plays the role of
the gravitational potential.

The relevant\footnote{``Relevant'' are termed those components of the
energy-momentum tensor that
do not depend on the low energy properties of the quasiparticles, {\it
i.e.} on the form and existence of a Lorentz metric.}\, 
components of the energy-momentum tensor are
($p_0= -E$, $p^0= E^{\rm com}$):
\begin{eqnarray}
\sqrt{-g} T^0{}_i & = & \sum_{\bf p} f {p}_i
= P_i\qquad \mbox{\sf Momentum density in either frame},
 \nonumber\\
\quad \sqrt{-g} T^0{}_0& = & -\sum_{\bf p} f E\qquad
\mbox{\sf (Negative) Laboratory frame energy density},
\nonumber\\
\sqrt{-g} T^k{}_i & = &  \sum_{\bf p} f p_i v_g^k \qquad
\mbox{\sf Laboratory frame momentum flux}, \nonumber\\
\sqrt{-g} T^i{}_0 & = & 
= -\sum_{\bf p} f E v_g^i\qquad  \mbox{\sf (Negative) Laboratory frame
energy flux}, \nonumber\\
 \sqrt{-g} T^{00} & = & \sum_{\bf p} f  p^0 = \sum_{\bf p} f  E^{\rm com}
 \qquad  \mbox{\sf Comoving frame energy density}.
\end{eqnarray}

We will further on keep the square root of the negative metric determinant for
structural clarity, though it is a space-time constant in our
considerations, as we assume the ``speed of light'' to be constant below.
The relation between the laboratory frame energy density   and
that in the comoving frame expresses the underlying Galilean invariance:
\begin{eqnarray}
-\sqrt{-g}T^0{}_0  & = &  \sum_{\bf p} f E =
\sqrt{-g} T^{00} +v_{\rm s}^i P_i\,. \label{labframeErel}
\end{eqnarray}

The additional hydrodynamic equation to be satisfied is that of the
conservation of the number of particles comprising the underlying liquid,
say, the number of $^3$He atoms. If temporal variations of the density are
negligible, this amounts to requiring that the total momentum density
\begin{equation}
{\bf j}= \rho{\bf v}_{\rm s} +{\bf P}
\end{equation}
has vanishing divergence, $\nabla\cdot {\bf j}=0$.
That we can neglect (temporal and spatial) variations of the density
 will be true in the case of $^3$He-A, where the ``speed of light''
$c_\perp$ is much smaller than the sound velocity.
This is in marked contrast to truly
``acoustic'' black holes, as they will exhibit large variations of the
density for velocities of the fluid above the speed of sound.

\subsection{Components of energy-momentum in local thermodynamic
equilibrium}
Local thermodynamic equilibrium is characterized by the local
temperature $T$ and local velocity ${\bf v}_n$ of the quasiparticle
system. The equilibrium distribution function is\cite{Khalatnikov}:
\begin{eqnarray}
f_{\cal T}=\frac 1
{1+\exp[T^{-1}(E({{\bf p}})-{\bf p}\cdot{\bf v}_n]}\,=
\frac 1
{1+\exp[T^{-1}(E^{\rm com}({{\bf p}})-p_i w^i(x))]}\,
\,.
\label{EquilibriumDistribution}
\end{eqnarray}
where
${\bf w}= {\bf v}_{\rm n} - {\bf v}_{\rm s}$ is the so-called counterflow
velocity.

In local thermodynamic equilibrium the components of energy-momentum
for the quasiparticle system (matter) are determined by the generic
thermodynamic potential (the pressure), which for fermions has the form
\begin{equation}
\Omega=-T\frac1{(2\pi\hbar)^D} \sum_{s}  \int d^Dp~{\rm ln}(1-f)\,.
\end{equation}
The momentum density is expressed in terms of $\Omega$ as
\begin{equation}
P_i=\sqrt{-g}T^0{}_{i}
=\frac1{(2\pi\hbar)^D} \sum_{s}  \int d^Dp~ f p_i= {\partial \Omega
\over \partial w^i}\equiv(\rho_{\rm n})_{ij}   w ^j \,.
\end{equation}
Here $\rho_n$ is the so-called
density of the normal component (in general a tensorial quantity),
so that the total current of the liquid:
${\bf j}=\rho {\bf v}_{\rm s} + \rho_{{\rm n}} ( {\bf v}_{\rm n} - {\bf
v}_{\rm s})\equiv
\rho_{{\rm s}}{\bf v}_{\rm s} + \rho_{{\rm n}} {\bf v}_{\rm n} $, where
$\rho_{\rm s}= \rho -\rho_{\rm n}$ is the density of the
superfluid component.

Considering below for simplicity the
(spatially) one-dimensional case ($D=1$), one has for the
 purely ``relativistic''
contribution such that $E^{\rm com}= |p_x|$ (and thus $(v^{\rm
com}_g)_x=1$):
\begin{equation}
\Omega= \sum_{s}\frac{\pi}{12\hbar}  T_{\rm eff}^2\,,\qquad
T_{\rm eff} = \frac{T}{\sqrt{1-w^2}}~\,,
\end{equation}
\begin{equation}
P_x=\sqrt{-g}T^0{}_x= \rho_{\rm n} w\,,\qquad\rho_{\rm n}=\frac
{2\Omega   }{1-w^2}\,.
\end{equation}
The other components are in the ``relativistic'' case then given as
\begin{equation}
T^{\mu\nu}  =    (\varepsilon  + \Omega )u^\mu
u^\nu+\Omega g^{\mu\nu}\,,\qquad \varepsilon=-\Omega +T{\partial \Omega
\over\partial T}=D\Omega\,,\qquad T^\mu{}_\mu=0~.
\label{QuasipStressTensorRel2}
\end{equation}
where $u^\alpha$ and
$u_\alpha=g_{\alpha\beta}u^{\beta}$, which satisfy the normalisation
equation $u_\alpha u^\alpha=-1$, are
\begin{equation}
u^0={1\over \sqrt{ 1 - w^2}}\,,\qquad u^i={v_{(n)}^i\over \sqrt{  1 -
w^2}}\,,\qquad u_i= { w_i \over \sqrt{{ 1 - w^2}}}\,,\qquad
u_0=-{1+{{\bf w} \cdot{\bf v}_{(s)}} \over
\sqrt{ { 1 - w^2}}}\,,
\label{4Velocity}
\end{equation}
The distribution of quasiparticles in local equilibrium in
(\ref{EquilibriumDistribution}) is expressed via the
temperature four-vector $\beta^\mu $:
\begin{equation}
f_{\cal T} = {1\over
1+\exp[-\beta^\mu p_\mu]}\,,\qquad \beta^\mu ={u^\mu\over  T_{\rm eff}}
=\left({1\over T}, {{\bf v}_n \over  T
}\right)\,,\qquad \beta^\mu\beta_\mu=-T_{\rm eff}^{-2}~.
\label{4Temperature}
\end{equation}
For the relativistic system, true equilibrium with vanishing entropy
production is established if $\beta^\mu$ is a timelike Killing vector:
\begin{equation}
\beta_{\mu;\nu}+
\beta_{\nu;\mu}=0~,\;\;{\rm or}\qquad
\beta^\alpha\partial_\alpha g_{\mu\nu}+
(g_{\mu\alpha}\partial_\nu +g_{\nu\alpha}\partial_\mu)\beta^\alpha=0
~.
\label{EquilibriumConditions}
\end{equation}
For a time independent, space dependent situation the  condition
$0=\beta_{0;0}=\beta^i\partial_i g_{00}$ gives
$\beta^i=0$, while the other conditions are satisfied when
$\beta^0={\rm constant}$. Hence true equilibrium requires that
${\bf v}_n=0$ in the frame of the velocity texture, and
$T={\rm constant}$. These are just the equilibrium conditions in
superfluids. They correspond to Tolman's law in general relativity,
with Tolman temperature being equal to the real temperature of the liquid:
\begin{equation}
T_{\rm eff}={T\over \sqrt{ 1 - v_{\rm s}^2}}={T\over \sqrt{-g_{00}}}\,.
\label{TolmanLaw}
\end{equation}

\section{Thermal states in the presence of a horizon}
\subsection{One-dimensional hydrodynamic equations}
To describe the thermal states in the presence of the horizon in a most
simple way, we consider a purely one-dimensional situation. This is, however, 
not completely unrealistic an assumption. 
Of course, we have to relax the condition
$j=$const., which does not allow for a velocity $v_{\rm s}$ gradient 
in the one-dimensional case. 
This, however, can be achieved by the inhomogeneous profile
of a tube. Then there is some flow in the transverse directions,
such that $\nabla\cdot{\bf j}=0$ is satisfied. An alternative possibility to
construct a one-dimensional horizon is to make $c_\perp(x)$  
space dependent\cite{grishated}. 
The mass conservation law  ${\bf j} = {\rm const.}$ also
implies that the nonzero quasiparticle momentum, which arises, say, due to
Hawking radiation, has to be compensated by a corresponding
change of the superflow velocity.
This is the mechanism of back-reaction of the superfluid
vacuum to quasiparticle dynamics. Back-reaction is, however, small
and can be neglected, namely
if the effective temperature of quasiparticles is
small compared to the superfluid transition temperature
$T_c$. This will always be true
in our approximation, for which $T< m_*c_\perp^2\ll T_c$.
Then, $P\ll \rho v_{\rm s}$.

We will investigate below a case for
which the quantum effects related to gravity,
including the  Hawking radiation process, can be neglected. This is permitted
if all the relevant energies are much higher than the Hawking
temperature:  $T\gg T_{\rm H}$, and $\hbar/\tau \gg T_{\rm H}$.
In the latter relation, $l=c\tau$ is the
mean free path, and the relation means that $l$ is small compared with the
characteristic length, within
which the velocity (the ``gravitational potential'')
changes: $l (\partial v_{\rm s}/\partial x) \ll 1$. This is just the condition
for the applicability of the two fluid hydrodynamic equations, where the
variables are the superfluid
velocity $v_{\rm s}$ which, when squared, plays the part of the
gravitational potential, as well as  temperature
$T(x)$ and velocity $v_{\rm n}(x)$ of the normal component, which
characterize the local equilibrium states of quasiparticles (matter).
The dissipative terms in the two fluid equations can
then be neglected in a zeroth order approximation,
since they are small compared to the reversibly hydrodynamic terms
by the above parameter $l (\partial v_{\rm s}/\partial x) \ll 1$.

If the superfluid velocity (gravity) field is fixed, the other hydrodynamic
variables,  temperature
$T(x)$ and velocity $v_{\rm n}(x)$ of ``matter'', are determined by the
conservation of energy and momentum. From equation (\ref{Econserv}) it
follows that the energy flux $Q$ carried by the quasiparticles is constant.
In the relativistic approximation one then has
\begin{equation}
Q=-\sqrt{-g} T^x{}_0
=  2\Omega\frac{v_{\rm n}(1 + wv_{\rm s})}{ { 1-w^2  }} ={\rm const} \,,
\qquad \Omega={(2s+1)\pi\over 12}T_{\rm eff}^2~.
\label{horizon1}
\end{equation}
From (\ref{Mconserv})  there results the first order
differential equation
 \begin{equation}
-\partial_x \left( 2\Omega~{v_{\rm n}w\over
1-w^2} +  \Omega\right )=    2\Omega {w\over
1-w^2}\partial_x v_{\rm s} ~ .
\label{horizon2}
\end{equation}

\subsection{Local equilibrium states with zero energy flux}
In the case that the energy flux is zero, equation (\ref{horizon1}) yields two
possible states.

\subsubsection{True equilibrium outside horizon}
Equation (\ref{horizon1}) is satisfied by the trivial solution
$v_{\rm n}=0$.  Then from (\ref{horizon2}) it follows that
$T={\rm constant}$. This corresponds to a true equilibrium state outside the
horizon. The effective temperature satisfies Tolman's law in
(\ref{TolmanLaw}): $T_{\rm eff}(x)= T/\sqrt{-g_{00}(x)}=
T/\sqrt{1-v_{\rm s}^2(x)}$. This equilibrium state cannot be continued across
the horizon: The effective temperature $T_{\rm eff}$ diverges when the horizon
is approached and becomes imaginary inside the horizon, where  $|w|=|v_{\rm s}|>1$.

\subsubsection{Thermal state behind  horizon}
Equation (\ref{horizon1}) is satisfied by $1+wv_{\rm s}=0$ (or $u_0=0$). Since
$w^2<1$, this solution can be valid only inside the horizon, where $v_{\rm
s}^2>1$. From (\ref{horizon2})  it follows that $\Omega={\rm
const}/(v_{\rm s}^2-1)$, and thus the temperature behaves as
$T^2 \propto \Omega (1-w^2)={\rm const}/ v_{\rm s}^2$,  or $T=T_{\rm
hor,left}/|v_{\rm s}|$, where $T_{\rm hor,left}$ is the temperature at the
horizon when approached from inside.   Thus inside the horizon one has
 a quasi-equilibrium state with inhomogeneous temperature. The effective
temperature behind the horizon follows a modified Tolman law:
\begin{equation}
T_{\rm eff}={T_{\rm hor,left}\over \sqrt{v_{\rm s}^2(x)-1}}\,.
\label{TolmannBehindHor}
\end{equation}
The above equation is valid only in 1+1 dimensions.
For larger spatial dimension $D>1$ one has $\Omega\propto
(v_{\rm s}^2-1)^{-(D+1)/2D}$ and $T_{\rm eff}\propto
(v_{\rm s}^2-1)^{-1/2D}$, so that
$T\propto (v_{\rm s}^2-1)^{(D-1)/2D}/|v_{\rm s}|$.

The thermal state behind the horizon is no longer determined by the
Hamiltonian $H(x>0)=-p_0=E$, but can be described as a local
equilibrium state if one uses the kinetic momentum as the Hamiltonian:
\begin{eqnarray}
f(p,x<0)= {1\over 1+\exp(H(x<0)/T_{\rm hor,left})}~, \\
~H(x<0)=-p^1\,,\qquad p^1=
(1-v_{\rm s}^2)p+v_{\rm s}E=p+v_{\rm s}|p|=-p^0\dot x~.
\label{DistributionBehindHor}
\end{eqnarray}
Here $\dot x=dE/dp$, so that
$p^1$ is a ``mass'' multiplied by a velocity and thus represents a kinetic
momentum.  The reason why the kinetic momentum plays the part of the
Hamiltonian is that
\begin{equation}
{p^1\over T_{\rm hor,left}}= \beta^\mu(x<0)p_\mu\,,\qquad
\beta^\mu=\left({v_{\rm
s}\over T_{\rm hor,left}}\,,\,
 {1-v_{\rm s}^2 \over T_{\rm hor,left}}\right)\,.
\label{ConKillingVector}
\end{equation}
Here $\beta^\mu(x<0)$ is a two-temperature,
which satisfies the CKV condition
$\beta_{\mu;\nu}+\beta_{\nu;\mu}\propto  g_{\mu\nu}$.
It can be verified by direct calculation that
\begin{equation}
\beta_{\mu;\nu}+
\beta_{\nu;\mu}= g_{\mu\nu}\partial_x(1-v_{\rm s}^2)\,.
\label{ConKillingCondition}
\end{equation}
For massless relativistic particles the CKV condition
is sufficient for thermal equilibrium states without dissipation to exist.
Thus, despite the
inhomogeneous temperature and nonzero normal velocity $v_{\rm n}$,
the state behind the horizon becomes a true equilibrium state in the
low energy relativistic limit.

Existence of nondissipative thermal states behind the horizon is a
consequence of the 1+1-dimensional situation, in which the
metric is conformally flat.
The coordinate transformation $\hat t=t -\int^x dx'
v_{\rm s}/(1-v_{\rm s}^2)~, ~\hat x=x$ transforms the vector
$\beta^\mu \rightarrow\hat\beta^\mu=(0,1-v_{\rm s}^2)$, so that $\hat
p^1=-i(1-v_{\rm s}^2)\partial_{\hat x}$. Another coordinate transformation
$dx/(1-v_{\rm s}^2)=d\hat x$ transforms $p^1$ to
$-i\partial_{\hat x}$. On the other hand, these two transformations
give the conformally flat metric $d{\hat s}^2
=(1-v_{\rm s}^2)(-dt^2 + dx^2)$
with the conformal factor $1-v_{\rm s}^2$. This implies that $p^1$ is a
conformal Killing vector satisfying the CKV condition in
(\ref{ConKillingCondition}).

\subsubsection{Matching thermal states across horizon}

Since the superluminal dispersion provides an energy
exchange between the matter (quasiparticles) inside and outside the
horizon,  the temperature must be continuous across the
horizon. Thus
$T_{\rm hor,left}=T_{\rm hor,right}$. It is important to note
that even in the
comoving frame, the blueshifted energies are finite and much smaller than the
``Planck'' scale, on which the hydrodynamic formalism breaks down. 
Consequently, the
temperature and the normal fluid velocity can be well determined not
only on both sides of the horizon but even at the horizon itself, where all
the characteristic scales are determined by the first nonlinear correction
to the quasiparticle spectrum. In our 1+1d case thermal states on
both sides of the horizon do, however,
 not depend on the details of this thin layer. 
The only effective output of the membrane is that it establishes the boundary
condition $T_{\rm hor,left}=T_{\rm hor,right}$. 

Thus, assuming that at $+\infty$ the superfluid velocity vanishes, one has
the following temperature profile on both sides of the horizon:
$T(x>0)=T_\infty$,
$T(x<0)=T_\infty/|v_{\rm s}(x)|$. The distribution functions on both
sides of the horizon have the form
\begin{eqnarray}
f(p,x>0)= {1\over 1+\exp(-p_0/T_\infty)} ~, \qquad p_0=|p| +v_{\rm s}p\,,\\
f(p,x<0)= {1\over 1+\exp(-p^1/T_\infty)}~,\qquad p^1=v_{\rm s}|p| +p\,.
\label{DistributionAcrossHor}
\end{eqnarray}
There is a symmetry between the states inside and outside the
horizon. While outside the horizon one has Tolman's law, $T(x>0)=T_{\rm
eff} \sqrt{1-v_{\rm s}^2}$,
the temperature behind the horizon is obtained from
Tolman's law by the substitution
$v_{\rm s}\rightarrow 1/v_{\rm s}$, {\it i.e.}, $T(x<0)=T_{\rm eff}
\sqrt{1-1/v_{\rm s}^2}$. The same procedure applies to $w$:
One has $w=-v_{\rm s}$ outside the horizon and
$w=-1/v_{\rm s}$
inside. The
temperature two-vector for the state behind horizon
\begin{equation}
\beta^\mu(x<0)=(\beta^0, \beta^1)= {1\over T_\infty}(-v_{\rm s},
1-v_{\rm s}^2)\,,\qquad\beta_\mu(x<0)=(\beta_0, \beta_1)={1\over T_\infty}
(0, 1)\,,
\label{betaInside}
\end{equation}
can be compared with the temperature two-vector for the true equilibrium
outside the horizon
\begin{equation}
\beta^\mu(x>0)={1\over T_\infty} (1, 0)\,   ,\qquad \beta_\mu(x>0)={1\over
T_\infty}(v_{\rm s}^2-1 ,-v_{\rm s})~.
\label{betaOutside}
\end{equation}
The relation between these two-temperatures shows that the states across
the horizon are dual to each other:
\begin{equation}
\beta^\mu(x>0)=\epsilon^{\mu\nu} \beta_\mu(x<0)~,
\label{betaRelation}
\end{equation}
{\it i.e.}, the space and time components of the two-temperature transform to
each other across the horizon ($\epsilon^{11}=\epsilon^{22}=0\,,\;
\epsilon^{12}=-\epsilon^{21}=1$).
It also appears that the thermal equilibrium behind the horizon is the
same as the thermal equilibrium outside the horizon, if the
superfluid velocity and the speed of light are interchanged. Behind the
horizon the speed of light plays the part of the superfluid velocity and
vice versa. This corresponds to the case of a  soliton moving in
superfluid $^3$He-A, where the speed of light changes in space, while the
superfluid velocity remains constant\cite{grishated}.

\subsection{Local equilibrium states with nonzero energy flux}

If the energy flux is  nonzero, we express
$\Omega(v_{\rm s},w)=Q(1-w^2)/2v_{\rm n}(1+wv_{\rm s})$ using
(\ref{horizon1}), and
substitute it into (\ref{horizon2}). One obtains an equation for
$w(v_{\rm s})$:
\begin{equation}
 {\partial w \over \partial v_{\rm s}}=  - \frac{1-w^2 } {1-v_{\rm s}^2 }~,
\label{w(v)equation}
\end{equation}
which has the solution
\begin{equation}
w=  \frac{(1-v_{\rm s})-C(1+v_{\rm s}) } {(1-v_{\rm s})+C(1+v_{\rm s})   }~,
\label{w(v)solution}
\end{equation}
where $C$ is a constant.
Using this equation we have the sequence of relations:
\begin{eqnarray}
v_{\rm n} & = & w+v_{\rm s}=  (1-C)\frac{1-v_{\rm s}^2 } {(1-v_{\rm
s})+C(1+v_{\rm s})
}\,,\nonumber\\ 1+wv_{\rm s} & = &(1+C)\frac{1-v_{\rm s}^2
} {(1-v_{\rm s})+C(1+v_{\rm s}) }\,,\nonumber\\
\frac{v_{\rm n}(1 + wv_{\rm s})}{ {1-w^2  }} & = &(1-v_{\rm s}^2)
{1-C^2\over 4C}~.
\label{vn}
\end{eqnarray}
Since $|w|\leq 1$, one has  $C_+>0$ outside the horizon,
where
$|v_{\rm s}|<1$, and
$C_-<0$ inside the horizon, where
$|v_{\rm s}|>1$.
The energy flux is
\begin{equation}
Q=    2 {(2s+1)\pi\over 12}  {1-C^2\over
4C}(1-v_{\rm s}^2) T_{\rm eff}^2~.
\label{Q}
\end{equation}
Since the energy flux can be only
inward, $Q<0$,  (in the absence of Hawking radiation), one has $C^2>1$.
From (\ref{Q}) it follows that the effective temperature exhibits
Tolman's law\footnote{The
relativistic Tolman law of a Lorentzian space-time obviously
involves corrections from the nonrelativistic tail.
These corrections are, however, small far enough away from the horizon.}\,,
determined by  the temperature at $+\infty$ (if
$v_{\rm s}(+\infty)=0$ is assumed):
\begin{equation}
  T_{\rm eff}={ T_\infty\over \sqrt{|1-v_{\rm s}^2| }}\,,\qquad
T^2_\infty = |Q|{24 \over
(2s+1)\pi}\frac{
 |C| }{ C^2-1 }~ .
\label{TolmanLaw2}
\end{equation}
It follows from the above relation that a modified form of Tolman's law,
telling us that the ``temperature'' detected by a local observer,
$T^{\rm eff}={\rm const.}  /\sqrt{|-g_{00}|} =
{\rm const.}/{\sqrt{|1-v_{\rm s}^2|}}$,
is fulfilled even in the absence of complete equilibrium.
``Temperature'' is set in quotation marks,
because $T^{\rm eff}$ represents only an effective ``relativistic''
temperature of the quasiparticles appearing in the distribution
function, while the true temperature of the superfluid still remains
$T$.
The effective temperature diverges at the horizon, while the true
temperature remains finite, but changes in space:
\begin{equation}
 T= \sqrt{1-w^2 }T_{\rm eff}=  T_\infty{ \sqrt{1-w^2 }\over
\sqrt{|1-v_{\rm s}^2| }}=
T_\infty  {2\sqrt{|C|} \over  |1-v_{\rm s}+ C(1+v_{\rm s})|}~,
\label{realT}
\end{equation}
and approaches at the horizon the value
\begin{equation}
 T_{\rm hor}=  T_\infty  \sqrt{|C|}  ~.
\label{realThor}
\end{equation}
Since the true temperature must be
continuous across the horizon, the condition $C_-=-C_+$ is enforced.

Hence the continuity of $T$ and the conservation of the energy flux completely
determine the thermal state inside the horizon in terms of the temperature
at positive infinity,
$T_\infty$, and the energy flux $Q$. In terms of $T_\infty$ and
$v_{\rm n}(+\infty)$, 
the relations $v_{\rm n}(+\infty)=(1-C_+)/(1+C_+)$ or
$C_+=(1-v_{\rm n}(+\infty))/(1+v_{\rm n}(+\infty))$ hold.
In the limit of vanishing flux,
$|C|\rightarrow 1$, one obtains the results of the previous subsection:
\begin{eqnarray}
{\rm outside~horizon}:~~~~  T=T_\infty~,~v_{\rm n}=0~,~w(x)=-v_{\rm s}(x)~,
\label{realTZeroFlux1}\\
{\rm inside~horizon}:~~~~ T(x) ={T_\infty\over
|v_{\rm s}(x)|}~,~v_{\rm n}(x) ={v_{\rm s}^2(x)-1\over v_{\rm s}(x)} ~,~w(x)
=-{1\over v_{\rm s}(x)}~.
\label{realTZeroFlux2}
\end{eqnarray}

\section{Properties of the horizon}

\subsection{Kinks at the horizon}
In the presence of superluminal dispersion, all physical quantities
are continuous at the horizon. On the other hand,
the thickness of the membrane, determined by the nonlinear
disperion in (\ref{CharacteristicLengthScale}) with $E=T_\infty$, is very
small and can be neglected. In this limit the thermal quantities
experience kinks. For example,
$\nabla  T|_{+0}=\nabla  T|_{-0} (C_+-1)/(C_--1)$;
$\nabla
w|_{\pm 0}=-C_\pm \nabla v_{\rm s}(x=0)$. In the case of zero flux, equations 
(\ref{realTZeroFlux1}-\ref{realTZeroFlux2}), the jump in the
derivative of the temperature is
$\nabla T|_{+0}-\nabla T|_{-0}=2\pi T_\infty T_{\rm H}$, and does not
depend on details of the dispersion. This means that in the limit of a
purely relativistic system, the presence of a nonzero temperature at
infinity implies a  singularity at the horizon. This  
singularity at the horizon cannot be removed, since in the presence of
``matter'' with nonzero temperature the system is not invariant under
coordinate transformations, and this produces a kink in temperature
at the horizon.

\subsection{Entropy related to horizon}

The thermal states also have an entropy profile, for which we can distinguish
the part coming from the bulk and that from the horizon.  Typically the entropy
related to the horizon is dominating, since it involves the ``Planck''
scale parameter. We again discuss the 1+1d case, where the horizon contribution
is only logarithmically larger than the contribution of the bulk.

The true entropy of superfluid, which is carried by its normal subsystem
at $T_\infty \gg T_{\rm H}$, is
\begin{equation}
 {\cal S}=\int d^D r~  S\,,\qquad S={\partial \Omega\over \partial T}~.
\label{TotalEntropy}
\end{equation}
The ``relativistic'' entropy, which is measured by a local observer living
in the quasiparticle world is the effective entropy
\begin{equation}
S_{\rm
eff}={\partial\Omega\over \partial T_{\rm
eff}} = {\partial\Omega\over \partial T} {\partial T\over \partial T_{\rm
eff}} =S\sqrt{1-w^2}~.
\label{EffectiveEntropy}
\end{equation}
In the presence of a horizon, the {\em total true entropy} can be divided into
3 contributions:
\begin{equation}
 {\cal S}={\cal S}_{\rm ext} + {\cal S}_{\rm int} +{\cal S}_{\rm
hor}~.
\label{TotalEntropyDivison}
\end{equation}
 The exterior entropy, which comes from the bulk liquid,
is proportional to the size $L_{\rm ext}$ of the external region:
${\cal S}_{\rm ext}
\propto T_\infty  L_{\rm ext}$. A similar estimate holds for the
entropy of the interior region, ${\cal S}_{\rm int}
\propto T_\infty  L_{\rm int}$. The entropy related to the
horizon stems from the logarithmically divergent
contribution at the horizon:
\begin{eqnarray}
{\cal S}_{\rm
hor} & = &{(2s+1)\pi \over 6} \left(\int_{-1/\kappa_{\rm s}}^{-x_0} +
\int_{x_0}^{1/\kappa_{\rm s}}\right)dx {T_\infty\over |1-v_{\rm s}^2|}\,,
\qquad 1+v_{\rm s}\approx\kappa_{\rm s} x\,,
\label{HorizonEntropy}
\end{eqnarray}
where we have used 
relations (\ref{realTZeroFlux1}) and (\ref{realTZeroFlux2}) (valid for
the limes of vanishing energy flux).
According to (\ref{3roots}), the ultraviolet cutoff $x_0$ is
provided by the thickness of the shell (membrane) determined by the nonlinear
dispersion, which gives (see equation 
(\ref{CharacteristicEnergyScale}) with $E=T_\infty$):
\begin{equation}
x_0\kappa_{\rm s}=
\tilde x_0\sim\tilde T_\infty^{2/3}=\left(T_\infty\over E_{\rm
Planck}\right)^{2/3} ~.
\label{ShellThickness}
\end{equation}
As a result one has
\begin{equation}
{\cal S}_{\rm
hor} ={(2s+1)  \over 18}   { T_\infty \over T_{\rm H}}
\ln \left({E_{\rm Planck}\over T_\infty}
\right)  ~ .
\label{HorizonEntropy2}
\end{equation}
This relation
also means that the density of quasiparticle states diverges logarithmically
at the horizon:
\begin{eqnarray}
N_{\rm hor}(E) & = &(2s+1)\int {dxdp\over 2\pi} \delta(E-E(p,x))={(2s+1)  \over
6\pi^2     T_{\rm H} }\ln
\left({E_{\rm Planck}\over E}\right)\,,
\nonumber\end{eqnarray}
\begin{equation}
T_{\rm H}\ll E\ll E_{\rm Planck}\,.
\label{DOS}
\end{equation}

\subsection{Dissipation at the horizon}
Behind the horizon the conditions for the true equilibrium in
(\ref{EquilibriumConditions}) are not satisfied,
$u_{\mu\nu}\equiv \beta_{\mu;\nu}+
\beta_{\nu;\mu}\neq 0$, cf. equation (\ref{ConKillingCondition}):
\begin{equation}
u_{00}=  {1\over T_\infty} (1-v_{\rm s}^2)\partial_x
(v_{\rm s}^2) \,,\qquad u_{01}={v_{\rm s}\over T_\infty}  \partial_x
(v_{\rm s}^2)\,,\qquad   u_{11}=-{1\over T_\infty}  \partial_x
(v_{\rm s}^2)          ~.
\label{ModifiedEquilibriumConditions}
\end{equation}
But quasi-equilibrium CKV conditions are met,
with the conformal function being equal to the local gravitational field,
normalized to the temperature at infinity:
\begin{equation}
u_{\mu\nu}=   2g_{\mu\nu}\Phi\,,\qquad \Phi = -{1\over T_{\rm \infty}}
\partial_x \left({g_{00}\over 2}\right)~.
\label{CKVconditions}
\end{equation}
Since the true temperature and the normal component velocity, $T$ and
$v_{\rm n}$, are space dependent, there is entropy production and
dissipation stemming from thermal conductivity, giving a term
$\frac{\kappa}T (\nabla T)^2$,
 and second viscosity, contributing a term
 $\zeta_2 ({\rm div}\, {\bf v}_{\rm n})^2$. In the
truly relativistic r\'egime with massless fermions dissipation is,
however, absent since the CKV conditions are
satisfied.  This fact is also known for the bosonic case,
superfluid $^4$He, where,
according to Khalatnikov\cite{Khalatnikov},
these two transport parameters are zero in a pure phonon gas.
They become nonzero for any deviation from the linear phonon spectrum.
Thus, if the Hawking process is neglected, in the purely
 relativistic r\'egime, there will be
global equilibrium states on both sides of the horizon, for any given
temperature at infinity.  Dissipation occurs only due to deviations
from the relativistic spectrum.
These deviations are most pronounced and effective within the thin shell
of the horizon vicinity, so that the main source of
dissipation in a 1+1d superfluid, in the presence of
a Landau horizon, will be concentrated at the membrane. The dissipation is
determined by the gradient of the superfluid velocity at the horizon.

This, then, is the thermal counterpart of Hawking
radiation. The latter becomes important only when the
quantum limit is approached, {\it i.e.}, when $T_\infty$ approaches the Hawking
temperature
$T_{\rm H}$. The hydrodynamic dissipation at $T_\infty\gg T_{\rm H}$ does the
same job as Hawking radiation at $T=0$, leading to the final extinction of
the black hole.  In our case, this means deceleration of the superfluid
below $|v_{\rm s}|=1$ and thus the consequent 
merging of black and white hole horizons (cf. Fig.\ref{BHWH}),   
and, finally, their mutual annihilation.

\subsection{Quantum corrections}

When the external temperature is low, and comparable to the Hawking
temperature $T_{\rm H}$, the quantum correction to the effective action
becomes important. The quantum part of the action, representing outside the
horizon the energy of the Boulware vacuum, is given by (see for
example equation (30) in Ref.\cite{Gusev}):
\begin{equation}
\Omega_{\rm quantum}= {2s+1\over 192\pi}\sqrt{-g} R{1\over \partial_\mu
g^{\mu\nu}\partial_\nu} R\,,
\label{QuantumAction}
\end{equation}
where $R$ is the Ricci scalar. If we are interested in the time independent 
action, then the metric $g_{\mu\nu}$ and the curvature $R$ are
time independent as well. 
Let us first consider the space outside the horizon, where we have
\begin{equation}
 \partial_\mu g^{\mu\nu}\partial_\nu =\partial_x(1-v_{\rm s}^2)\partial_x\,.
\label{metric}
\end{equation}
After an integration by parts, we obtain
\begin{equation}
\Omega_{\rm quantum}(x>0)= -{2s+1\over 192\pi}{1\over 1-v_{\rm s}^2}\sqrt{-g}
\left(\partial_x^{-1}R\right)\left(\partial_x^{-1}R\right)\,,
\label{QuantumAction2}
\end{equation}
The curvature $R$ calculated by using the ``acoustic'' metric outside the
horizon yields
\begin{equation}
R=\partial_x^2 (v_{\rm s}^2)\,,
\label{curvature}
\end{equation}
and thus the quantum correction becomes
\begin{equation}
\Omega_{\rm quantum}(x>0)=-{2s+1\over 192\pi} (\partial_x  (v_{\rm s}^2))^2{1\over
1-v_{\rm s}^2}\,.
\label{QuantumActionOutside}
\end{equation}
This coincides with the result of Ref.\cite{BrickWalls}, obtained outside
the horizon for the Boulware vacuum. Like in the case of the thermal
energy, we may extend this result to the region  behind the horizon.
Behind the horizon, one has to consider the vacuum state as being the
limit of vanishing normal component contribution (the heat bath),    
which keeps the value of the normal component velocity $v_n$ fixed. 
This means that we have to consider the
vacuum state with respect to the moving heat bath. As a result, the
coordinate dependence of the quantum action behind the horizon is
determined by the counterflow velocity $w$, instead of the superfluid
velocity $v_{\rm s}$. Since $w=-1/v_{\rm s}$ (in the limit of vanishing energy flux), 
the result is given by (\ref{QuantumActionOutside}), 
using the substitution $v_{\rm s}^2\rightarrow 1/v_{\rm s}^2$:
\begin{equation}
\Omega_{\rm quantum}(x<0)=-{2s+1\over 192\pi} (\partial_x  (v_{\rm s}^2))^2{1\over
  v_{\rm s}^2(v_{\rm s}^2-1)}\,.
\label{QuantumActionInside}
\end{equation}
Close to the horizon both regions, outside and inside the horizon,
therefore contribute in a similar way:
\begin{equation}
\Omega_{\rm quantum}\approx -{(2s+1)\pi\over 12} {T_{\rm H}^2\over
|1-v_{\rm s}^2|}~.
\label{QuantumAction3}
\end{equation}
The total energy density close to the horizon, quantum + thermal, 
is given by
\begin{equation}
\Omega({\rm total}) =\Omega({\rm quantum})+\Omega(T)\approx
{(2s+1)\pi\over 12} {T_{\infty}^2-T_{\rm H}^2\over |1-v_{\rm s}^2|}~,
\label{TotalAction}
\end{equation}
In agreement with more general theory (see, {\it e.g.},\cite{BrickWalls}), 
the diverging terms in the thermal and quantum 
energy compensate each other if the external temperature 
at spatial infinity equals the Hawking temperature,
$T_{\infty}=T_{\rm H}$, that is, in the Hartle-Hawking state (the
stress-energy tensor in the Hartle-Hawking state, defined by the 
vacuum of the freely falling observer, is regular at the
horizon).  However, despite such a compensation, the Hartle-Hawking state
remains dissipative: The true temperature $T$ is inhomogeneous behind the
horizon and the velocity $v_n$ of the normal component is nonzero. 
This leads to dissipation behind the horizon, due to the high frequency
nonlinear dispersion.

The quantum correction does not change the entropy of the horizon in the
thermal state of (\ref{HorizonEntropy2}). But in the particular state
of $T_{\infty}=T_{\rm H}$, the Hartle-Hawking state, the quantum
correction and the entropy of the horizon are related by a general
theorem. According
to Refs.\cite{IyerWald,JacobsonKangMyers,FrolovFursaev},
 this entropy is expressed in terms of the Lagrangian as
\begin{equation}
S_{\rm hor}=-8\pi\int_{\Sigma}t_\mu n_\nu t_\lambda n_\rho{\partial L\over \partial
R_{\mu\nu\lambda\rho}}d\sigma~,
\label{Entropy}
\end{equation}
In the $1+1$-dimensional case, for which 
$R=2R_{0101}/g= -2R_{0101}$, it follows that
\begin{equation}
S_{\rm hor}=-8\pi{\partial \Omega\over \partial
R_{0101}}{\bigg|}_{\rm hor}= 16\pi{\partial \Omega\over \partial
R}{\bigg|}_{\rm hor} ~.
\label{EntropyOneD}
\end{equation}
Using (\ref{QuantumAction}) for the Lagrangian, one obtains
\begin{eqnarray}
S_{\rm hor}& = &{(2s+1)\over 12} \left(\partial_x^{-1}\left({1\over
|1-v_{\rm s}^2|} \partial_x^{-1}R\right)\right)_{\rm hor}\nonumber\\
& = & -{(2s+1)\over 12}
{\left(
\ln | 1-v_{\rm s}^2|\right)_{\rm hor}}={(2s+1)\over 18} \ln {E_{\rm
Planck}\over T_{\rm H}} \,,
\label{Entropy2}
\end{eqnarray}
where we took into account the cutoff induced by the high-energy
dispersion,
$|1-v_{\rm s}^2|\sim  (T_{\rm H}/E_{\rm Planck})^{2/3}$.
The expression (\ref{Entropy2})  coincides with the
estimation of the entropy of the horizon in equation 
(\ref{HorizonEntropy2}), if
$T_{\infty}=T_{\rm H}$, {\it i.e.}, for the Hartle-Hawking state.

\section{Concluding Remarks}

Horizons and ergoregions can be simulated in a multitude of
different systems (cf. the overview in\cite{JacobsonTalk}).
The most promising systems,
displaying quantum behaviour related to a nontrivial effective
space-time, are superfluid $^4\!$He, superfluid $^3\!$He,
and dilute Bose-Einstein condensates\cite{BoseCondensate,HoleForLight},
which can be conveniently manipulated by lasers\cite{BECExpReviews}.

The Hawking-analogous process itself is exceedingly difficult to
measure in experimental practice. The Hawking temperature given by
$2\pi T_{\rm H} = \hbar\kappa_{\rm s}$ is, in experimentally  relevant units,
$T_{\rm H} = 1.22\, 
{\rm nK}\times \kappa_{\rm s} [10^3{\rm sec}^{-1}]$\,. In $^3\!$He-A, where
the velocity gradient attainable, in principle, is limited by
$\kappa_{\rm s} < c_\perp /\xi_A $, this implies that $T_{\rm H} < 0.1\, \mu$K. In
superfluid $^4\!$He, the corresponding limitation, $\kappa_{\rm s} < c/\xi$,
for $T_{\rm H}$ is much less severe, and comparatively
large $T_{\rm H}$ would be available ($T_{\rm H}\sim O(10\, {\rm mK})
\sim O(10^{-2} T_c))$,
for $\kappa_{\rm s}\sim O(10^{-2} c/\xi$).
The Hawking process would, however, probably 
be masked by non-Hawkingian
density perturbation excitations, generated by trying to force this dense
fluid to move with large gradient of velocities near the speed of sound.
Matters will possibly improve
for a dilute Bose-Einstein condensate, {\it e.g.} a sodium 
vapour\cite{ketterlecriticalVc}, 
$T_{\rm H}\sim O(0.1\, {\rm nK})\sim O(10^{-4}\,
T_c)$, for $\kappa_{\rm s}\sim O(10^{-2} c/\xi)$, because there to sustain this large
flow gradient without generating other perturbations not related to the 
Hawking process might be simpler.

It appears, however, that the classical dissipative signature of a horizon
for temperatures well above $T_{\rm H}$, which we have been discussing in
section 5.3, will more straightforwardly be detected
 with current technological means. The results of the present 
work can be extended
to the more realistic  2+1 and 3+1 cases. In all cases the thin
shell, which includes both sides of the horizon, is of great importance. In the
shell the large blueshift is compensated by the first nonlinear
correction to the linear ``relativistic'' energy spectrum.  The thickness of
the shell is much smaller than the scale of the velocity profile (the radius
of the horizon in the black hole analogy), but also much larger than the
microscopic length scale (the Planck length in the analogy). This allows for 
a determination of 
the temperature of the thermal state everywhere across the
horizon. This shell is responsible for the most important properties of the
thermal states in the presence of the horizon, including the entropy and
entropy production, both concentrated in or close to the shell.

In principle even the 1+1-dimensional case discussed here
can be  realized. The condition for the applicability of our
1+1-dimensional results is that the transverse energy level separation for,
say, a torus geometry  of the BEC be larger than the temperature, such that
these levels are not (appreciably) occupied. Such a torus trap may be 
realized by using a highly anisotropic trap, whose trapping frequency   is
``soft'' in only one, ring-shaped, direction.  The condition  on 
the transverse energy level separation reads
$T\ll \hbar c/2d$, where 
$d$ is the diameter of the torus of cross section $\sigma=\pi d^2/4$.
For that same sodium BEC we have been using for an  estimation above, for
$c/2d\sim 10^3\,$sec$^{-1}$ ($c\sim O(1-10\,\, $mm/sec), $d\sim O( \mu$m)),
it turns out that $T\ll 10\, {\rm nK} \sim O(0.01\cdots 0.1\, T_c)$ 
is to be fulfilled, which appears feasible (recently improved 
calorimetry has already made it possible\cite{ketterlehydrodynamics} 
to measure temperature changes of less than 10 nK). 
A closer investigation of this classical
dissipative process in these systems
should, under general conditions, then
take into account 
the effect of back-reaction on the phenomenon of horizon entropy production.


\nonumsection{Acknowledgements}
\noindent
G. E. Volovik 
thanks Valeri Frolov and Alexei Starobinsky for helpful discussions.
His work was supported by the Russian Foundation for Fundamental
Research Grant No. 00-15-96699, 
by the Intas grant and by the European Science Foundation.
U. R. Fischer was supported by
the Human Capital and Mobility Programme of the European Union under
Contract No. ERB FMGE CT980122 and by the DFG (FI 690/1-1).

\nonumsection{References}


\vspace*{-1em}
\noindent
\begin{thebibliography}{499}
\bibitem{hawkingnature} S. W. Hawking, ``{Black hole
explosions?}'', Nature {\small \bf 248}, 30-31 (1974).

\bibitem{vissersonic}
M. Visser, ``{Acoustic black holes: horizons,
ergospheres, and Hawking radiation}'', Class. Quantum Grav. {\small \bf 15},
1767-1791 (1998); M. Visser,  
``{Hawking radiation without black hole entropy}'', 
Phys. Rev. Lett. {\small \bf 80}, 3436 (1998). 

\bibitem{kangted} T. Jacobson and G. Kang, ``{Conformal invariance
of black hole temperature}'', Class. Quantum Grav. {\small \bf 10}, L201-L206
(1993).

\bibitem{unruh1} W. G. Unruh, ``{Experimental Black-Hole
Evaporation?}'', Phys. Rev. Lett. {\small \bf 46}, 1351 (1981).

\bibitem{unruh2} W. G. Unruh, ``{Sonic analogue of black holes and
the effects of high frequencies on black hole evaporation}'', Phys. Rev.
D {\small \bf 51}, 2827-2838 (1995).

\bibitem{BHdispers} S. Corley and T. Jacobson, ``{Hawking spectrum
and high frequency dispersion}'', Phys. Rev. D {\small \bf 54}, 1568-1586
(1996).

\bibitem{Corley} S. Corley, ``{Computing the spectrum of black hole radiation
in the presence of high frequency dispersion: An analytical approach}'',
Phys. Rev. D {\small \bf 57}, 6280-6291  (1998).

\bibitem{BHlaser} S. Corley and T. Jacobson, ``{Black hole
lasers}'', Phys. Rev. D {\small \bf 59}, 124011 (1999).

\bibitem{origin} T. Jacobson, ``{On the origin of the outgoing
black hole modes}'', Phys. Rev. D {\small \bf 53}, 7082-7088 (1996).

\bibitem{grishated} T. A. Jacobson and G. E.  Volovik, 
``{Event horizons and ergoregions in $^3\!$He}'',
Phys. Rev. D {\small \bf 58}, 064021 (1998); T. A. Jacobson and G. E.  Volovik, 
``{Effective spacetime and Hawking radiation from a moving domain wall
in a thin film of $^3\!$He-A}'', JETP Lett. {\small \bf 68}, 874-880 (1998)
[Pis'ma Zh. \'Eksp. Teor. Fiz.. {\small \bf 68}, 833-838 (1998)].

\bibitem{parallel} G.E. Volovik, ``{Field theory in superfluid $^3\!$He:
What are the lessons for particle physics, gravity and high-temperature
superconductivity?}'', 
Proc. Natl. Acad. Sci. USA {\small \bf 96},
6042 - 6047 (1999);  G.E. Volovik, ``{$^3\!$He and Universe parallelism}'',
in ``Topological Defects and the Non-Equilibrium Dynamics of Symmetry
Breaking Phase Transitions'', Yu. M. Bunkov, H. Godfrin (Eds.),
pp. 353-387 (Kluwer, 2000). 

\bibitem{Khalatnikov} I. M. Khalatnikov,  ``{An Introduction to the Theory of
Superfluidity}'' (Benjamin, New York, 1965).

\bibitem{sakharov} A. D. Sakharov,  ``{Vacuum Quantum Fluctuations in Curved
Space and the Theory of Gravitation}'', Dokl. Akad. Nauk {\small \bf 177}, 70-71
(1967) [Sov. Phys. Dokl. {\small \bf 12}, 1040-41 (1968)].

\bibitem{grishapainleve} G. E. Volovik, ``{Simulation of
Painlev\'e-Gullstrand Black Hole in thin $^3\!$He-A film}'',
JETP Lett. {\small \bf 69}, 705-713 (1999)
[Pis'ma Zh. \'Eksp. Teor. Fiz.. {\small \bf 69}, 662-668 (1999)].

\bibitem{Tolman} R. C. Tolman, ``{Relativity, Thermodynamics and Cosmology}'' 
(Clarendon Press, Oxford, 1934).

\bibitem{Zimdahl} W. Zimdahl, ``{Cosmological particle production
and generalized thermodynamic equilibrium}'', Phys. Rev. D {\small \bf 57}, 2245-2254
(1998).

\bibitem{Martel} K. Martel and E. Poisson, ``{Regular coordinate systems for
Schwarzschild and other spherical spacetimes}'', gr-qc/0001069.

\bibitem{BoseCondensate}     L. J. Garay, J. R. Anglin, J.
I. Cirac and P. Zoller,    
``{Sonic analog of gravitational black holes in Bose-Einstein condensates}''
Phys. Rev. Lett. {\small \bf 85}, 4643-4647 (2000).  

\bibitem{Rice} M. Rice, ``{Superconductivity: An analogue of superfluid
$^3\!$He}'', Nature  {\small \bf 396}, 627-629
(1998); 
H. Mukuda 
{\small \it et al.}, ``{Spin-triplet
superconductivity in Sr$_2$RuO$_4$ identified by $^{17}\!$O Knight shift}'',
Nature  {\small \bf 396}, 658-660 (1998).

\bibitem{andreev} A. F. Andreev, ``{The thermal conductivity of the
intermediate state in superconductors}'', JETP {\small \bf 19}, 1228-31 (1964)
[Zh. \'Eksp. Th. Fiz. {\small \bf 46}, 1823-28 (1964)].

\bibitem{EMomtensor} G. E. Volovik, ``{Energy-momentum tensor of
quasiparticles in the effective gravity in superfluids}'', gr-qc/9809081.

\bibitem{LandauLifshitz2} L. D. Landau and E. M. Lifshitz, ``{The Classical 
Theory of Fields}'' (Pergamon Press, Revised Second Edition, 1962).  

\bibitem{Gusev}   Yu. V. Gusev and A. I. Zelnikov,  
``{Two-dimensional effective 
action for matter fields coupled to the dilaton}'', 
Phys. Rev. D {\small \bf 61}, 084010 (2000).

\bibitem{BrickWalls} S. Mukohyama and W. Israel, ``{ 
Black holes, brick walls and the Boulware state}'', Phys. Rev. D 
{\small \bf 58}, 104005 (1998).

\bibitem{IyerWald} V. Iyer and R. M. Wald, 
``{Some properties of the Noether charge
and a proposal for dynamical black hole entropy}'', 
Phys. Rev. D {\small \bf 50}, 846-864 (1994).

\bibitem{JacobsonKangMyers} T. Jacobson, G. Kang and R. C. Myers,  
``{On black hole entropy}'', Phys. Rev. D {\small \bf 49},
6587-6598  (1994).

\bibitem{FrolovFursaev} V. Frolov and D. Fursaev, ``{Black Hole Entropy in
Induced Gravity: Reduction to 2D Quantum Field Theory on the Horizon}'',
Phys. Rev. D {\small \bf 58},  124009 (1998).

\bibitem{JacobsonTalk}   T. A.  Jacobson, 
``{Trans-Planckian redshifts and the substance of the space-time river}'',   
Prog. Theor. Phys. Suppl. {\small \bf 136}, 1-17 (1999).  


\bibitem{HoleForLight}
U. Leonhardt and P. Piwnicki, ``{Relativistic Effects of Light in Moving
Media with Extremely Low Group Velocity}'', Phys. Rev. Lett. {\small \bf 84},
822-825 (2000); also cf. a Comment on this paper by
M. Visser, Phys. Rev. Lett. {\small \bf 85}, 5252 (2000), 
Reply {\small \it ibid.}
{\small \bf 85}, 5253 (2000). 


\bibitem{BECExpReviews} E. A. Cornell, J. R. Ensher and C. E. Wieman,  
``{Experiments in Dilute Atomic Bose-Einstein Condensates}'', cond-mat/9903109;
W. Ketterle, D. S. Durfee and D. M. Stamper-Kurn, ``{Making, probing and
understanding Bose-Einstein condensates}'', cond-mat/9904034. 

\bibitem{ketterlecriticalVc} C. Raman {\small \it et al.}, ``{Evidence for
a Critical Velocity in a Bose-Einstein Condensed Gas}'',
Phys. Rev. Lett. {\small \bf 83}, 2502 (1999). 

\bibitem{ketterlehydrodynamics} R. Onofrio {\small \it et al.},  
``{Observation of Superfluid Flow in a Bose-Einstein Condensed Gas}'', 
Phys. Rev. Lett. {\small \bf 85}, 2228 (2000).  

\end{thebibliography}
\end{document}